\documentstyle[aps,prc,graphicx,floats]{revtex}

\newcommand{\N}{N_{\rm evts}}
\newcommand{\Np}[0]{N'}
\newcommand{\mean}[1]{\left\langle #1 \right\rangle} 
\newcommand{\sample}[1]{\left\{ #1 \right\}} 
\newcommand{\cumul}[1]{\left\langle\!\!\left\langle #1 
  \right\rangle\!\!\right\rangle} 

\begin{document}

\preprint{ULB-TH-01/011}
\draft

\title{Flow analysis from multiparticle azimuthal correlations}

\author{Nicolas Borghini,$^{1}$
Phuong Mai Dinh,$^2$ 
and Jean-Yves Ollitrault$^{2,3}$}

\address{$^1$~Service de Physique Th\'eorique, CP225, 
Universit\'e Libre de Bruxelles, B-1050 Brussels}
\address{$^2$~Service de Physique Th\'eorique, CEA-Saclay, 
F-91191 Gif-sur-Yvette cedex}
\address{$^3$~L.P.N.H.E., Universit\'e Pierre et Marie Curie, 
4 place Jussieu,
F-75252 Paris cedex 05}

\maketitle

\begin{abstract}
We present a new method for analyzing directed and elliptic flow
in heavy ion collisions. 
Unlike standard methods, it separates the contribution of flow to 
azimuthal correlations from contributions due to other effects. 
The separation relies on a cumulant expansion of multiparticle 
azimuthal correlations, and includes corrections for detector 
inefficiencies. 
This new method allows the measurement of the flow of identified 
particles in narrow phase-space regions, and can be used in every 
regime, from intermediate to ultrarelativistic energies. 
\end{abstract}


\section{Introduction}
\label{s:intro}

In noncentral heavy ion collisions, it is possible to measure 
azimuthal distributions of outgoing particles with respect to the reaction 
plane. This is the so-called ``flow analysis'', which is 
being actively studied over a wide range of colliding energies, 
from below 25~MeV per nucleon in the center-of-mass 
system~\cite{Andronic:2001cx} to above 60~GeV~\cite{Ackermann:2001tr}.
The main motivation for such studies is that anisotropies in the 
azimuthal distributions 
are likely to contain much information on the
physics in the hot, dense central region of the collision
(see \cite{Reisdorf:1997fx,Herrmann:1999wu,Ollitrault:1998vz} for 
reviews). 
In particular, they may provide a signature of the formation 
of a quark-gluon plasma at ultrarelativistic energies 
\cite{Sorge:1999mk,Teaney:2000cw}.
Azimuthal distributions may also be of interest in connection with more 
exotic phenomena, such as the formation of disoriented chiral 
condensates \cite{Nandi:1999vb,Asakawa:2000dr}, or the study of 
parity and/or time-reversal violation\cite{Voloshin:2000xf}.
Finally, the combination of flow analysis and two-particle
interferometry yields a three-dimensional picture of the emitting 
source \cite{Voloshin:1996mc,Wiedemann:1998cr,Heiselberg:1999ik}, 
as was shown recently at the Brookhaven Alternating Gradient 
Synchrotron (AGS) \cite{Lisa:2000xj}.

In this paper, we propose a new method to measure azimuthal 
distributions. As usual, they will be characterized by 
their Fourier coefficients \cite{Voloshin:1996mz}
\begin{equation}
\label{defvn}
v_n\equiv\langle e^{in(\phi-\Phi_R)}\rangle
=\langle \cos n(\phi-\Phi_R)\rangle
\end{equation}
where $\phi$ is the azimuthal angle of an outgoing particle
measured in the laboratory coordinate system, 
$\Phi_R$ is the azimuth of the impact parameter (or reaction 
plane) and angular brackets denote a statistical average, 
over many events. 
The first two coefficients $v_1$ and $v_2$ are usually referred
to as directed flow and elliptic flow, respectively. 
The $v_n$'s can be measured for various particle species, as 
a function of transverse momentum and/or rapidity: we refer to 
these detailed measurements as to ``differential'' flow, 
following~\cite{Li:1999bh}. 
In this paper, we also discuss global measurements of $v_n$, 
averaged over a large phase space region,
typically corresponding to the acceptance of a detector. 
We call this ``integrated'' flow.  

Since the reaction plane $\Phi_R$ cannot be measured directly, 
the only way to obtain the coefficients $v_n$ experimentally is to deduce 
them from the azimuthal correlations between the outgoing particles:
the correlation between every particle and the reaction plane 
induces correlations among the particles (which we call hereafter
the ``flow correlations''), from which $v_n$ can be reconstructed. 
The method we propose here is based on a systematic analysis 
of multiparticle azimuthal correlations. 

The most widely used method for the flow analysis is that 
initially proposed by Danielewicz and Odyniec
\cite{Danielewicz:1985hn} (see also 
\cite{Poskanzer:1998yz,Ollitrault:1993ba,Ollitrault:1997di} for 
further developments), which relies on azimuthal correlations 
between two ``subevents''. 
It has recently been applied at intermediate energies 
in Darmstadt \cite{Andronic:2001cx,Taranenko:1999yh}, 
and at higher energies in 
Dubna \cite{Chkhaidze:2000gs,AbdAllah:2000qr,Simic:2001aq},
at the Brookhaven AGS \cite{Jain:1995cm,Chung:2000ny,Barrette:2001cb},
at the CERN Super Proton Synchrotron (SPS) 
\cite{Appelshauser:1998dg,Schlagheck:1999ja}
and, finally, at the Brookhaven Relativistic Heavy Ion Collider 
(RHIC) \cite{Ackermann:2001tr}.
An alternative, simpler method extracts flow from two-particle 
correlations \cite{Wang:1991qh} and is still in use, both at 
intermediate \cite{Prendergast:2000fv} and at ultrarelativistic 
energies \cite{Singh:1994ti,Lacey:2001}. 
Both methods are more or less equivalent: correlating two subevents 
amounts to summing two-particle correlations. 
In these analyses, one usually assumes that the only sources of 
azimuthal correlations are flow and, when necessary, 
transverse momentum conservation \cite{Danielewicz:1988in}.

However, we have shown in two papers 
\cite{Dinh:2000mn,Borghini:2000cm} that 
this assumption is no longer valid at SPS energies, 
where ``direct'', nonflow two-particle correlations become 
of the same magnitude as the correlations due to flow. 
Even when nonflow correlations are {\em a priori} smaller than flow 
correlations, they must be taken into account 
in order to obtain accurate and reliable results. 
Some sources of nonflow correlations are well known. 
One can attempt to avoid them experimentally by 
appropriate cuts in phase space \cite{Ackermann:2001tr}, 
or one can take them into account in the analysis 
\cite{Poskanzer:1998yz}, 
as was done for transverse momentum conservation 
\cite{Danielewicz:1988in,Borghini:2000cm}, 
for correlations from $\pi_0\to\gamma\gamma$ decays 
\cite{Aggarwal:1997iu,Raniwala:2000yz}, and for quantum correlations 
\cite{Dinh:2000mn}.  
But there is no systematic way to separate the effects 
of flow from other effects at the level of 
two-particle correlations. 

There have been several attempts in the past to go beyond 
two-particle correlations: analyses of the global 
event shape\cite{Danielewicz:1983we} allowed the first 
observations of collective flow at intermediate\cite{Gustafsson:1984ka}
and ultrarelativistic energies\cite{Ollitrault:1992bk,Barrette:1994xr},
which were not biased by nonflow correlations. 
Multiparticle azimuthal correlations were also used 
in \cite{Jiang:1992bw}. 
These methods are now considered obsolete because 
they apply only to the ``integrated'' flow, as defined above, 
whereas most of the recent analyses concentrate on the differential 
flow of identified particles, in particular kaons 
\cite{Chung:2000ny,Crochet:2000fz},
$\eta$ mesons \cite{Taranenko:1999yh}, 
$\Lambda$ hyperons \cite{Barrette:2001cb,Chung:2001je}, 
and antiprotons \cite{Barrette:2000ex}. 

In recent papers \cite{Borghini:2001sa,Borghini:2000iy},
we have shown for the first time 
that nonflow correlations can be removed systematically not only
for integrated flow, but also in analyses of differential flow. 
However, a limitation of this method when measuring $v_n$ 
is the interference with higher harmonics ($v_{2n}$, $v_{3n}$, etc.).
This interference may hinder the measurement of directed flow 
when elliptic flow is larger, which is likely to be the case 
at RHIC energies~\cite{Snellings:2000bt}. 

Here we present an improvement of this method, which is free from 
this limitation, and in many respects simpler. 
In particular, it no longer involves the event flow 
vector on which most analyses are based
\cite{Danielewicz:1985hn,Poskanzer:1998yz,Borghini:2001sa}.
As in our previous method, we perform a cumulant expansion 
of multiparticle azimuthal correlations, which eliminates 
order by order nonflow correlations, and can be used even if the 
detector does not have full azimuthal coverage.

In Sec.\ \ref{s:integrated}, we show how to construct the 
cumulants of multiparticle azimuthal correlations by means 
of a generating function. These cumulants allow us to 
reconstruct the integrated flow from the measured correlations. 
The method is extended to differential flow in Sec.\ \ref{s:differential}. 
The relation with other methods is discussed in 
Sec.\ \ref{s:previousmethod}. 
Results of Monte-Carlo simulations are presented in 
Sec.\ \ref{s:mc}. 
The most technical points are left to appendices: 
the construction of cumulants is explained in detail in 
Appendix \ref{s:cumulants};  
interpolation formulas used to obtain the 
cumulants from the generating function are given in 
Appendix \ref{s:interpolation}; 
acceptance corrections, which extend the validity of the method to 
detectors with partial azimuthal coverage, are derived in 
Appendix \ref{s:acceptance}; 
finally, statistical errors on the flow values deduced from the cumulants 
are evaluated in Appendix \ref{s:statistical}. 

The essential improvement on our previous method is the use of a
new generating function, defined in Sec.\ \ref{s:int_gene}, which 
corrects the limitations encountered in \cite{Borghini:2001sa}. 
These improvements are discussed in detail in Sec.\ \ref{s:previousmethod} 
and in Appendix \ref{s:cumulants};
they are seen clearly in the simulations presented in Sec.\ \ref{s:mc}.
In addition, the detailed discussions of acceptance corrections
(Appendix \ref{s:acceptance}) and statistical errors 
(Appendix \ref{s:statistical}) are completely new, although they 
also apply to our previous method. 
Apart from these differences, most of the material discussed in 
Secs.\ \ref{s:integrated} and \ref{s:differential} can be found 
in \cite{Borghini:2001sa}, although the present derivation is 
more transparent.

\section{Integrated flow}
\label{s:integrated}

In Sec.\ \ref{s:int_cumul}, we illustrate with a few examples 
the principle of the cumulant expansion of multiparticle 
azimuthal correlations, and show how it can be used to perform flow 
measurements with a better sensitivity than the previous methods.
Then we explain, in Sec.\ \ref{s:int_gene}, how to perform 
this expansion in practice, by means of a generating function.
In Sec.\ \ref{s:int_flow}, we derive the relations between the 
cumulants and the flow $v_n$, integrated over some phase space region.  
Using cumulants to various orders, one thus obtains different 
estimates for $v_n$. 
The uncertainties associated with each estimate due to nonflow 
correlations and limited statistics, and the resulting 
optimal choice, are examined in Sec.\ \ref{s:int_stat}.  
Finally, we discuss in Sec.\ \ref{s:weights} the generalization 
of the previous subsections to different, optimal particle weights.

\subsection{Cumulants of multiparticle azimuthal correlations}
\label{s:int_cumul}

We denote by $\phi_j$, with $j=1,\ldots ,M$, 
the azimuthal angles of the particles seen in an event 
with multiplicity $M$, 
measured with respect to a fixed direction in the detector 
(this was denoted by $\bar\phi_j$ in \cite{Borghini:2001sa}). 
In this paper, we shall be concerned with multiparticle 
azimuthal correlations, which we write generally in the form 
$\mean{\exp[in(\phi_1+\cdots+\phi_k-\phi_{k+1}-\cdots-\phi_{k+l})]}$, 
where $n$ is the Fourier harmonic under study
($n=1$ for directed flow, $n=2$ for elliptic flow), and 
the brackets indicate an average which is performed in two steps: 
first, one averages over all possible combinations 
of $k+l$ particles detected in the same event; 
then, one averages over all events. 

Correlations between $k+l$ particles can be generally decomposed into 
a sum of terms involving correlations between a smaller number of 
particles. 
Consider for instance the measured two-particle azimuthal correlation 
$\mean{e^{in(\phi_1-\phi_2)}}$. It can be written as 
\begin{equation}
\label{defc2}
\mean{e^{in(\phi_1-\phi_2)}}=
\mean{e^{in\phi_1}}\mean{e^{-in\phi_2}} +
\cumul{e^{in(\phi_1-\phi_2)}}, 
\end{equation}
where $\cumul{e^{in(\phi_1-\phi_2)}}$ is by definition the second order 
cumulant. 
To understand the physical meaning of this quantity, 
we first consider a detector whose acceptance is isotropic, i.e., 
which does not depend on $\phi$. Such a detector will be called a 
``perfect'' detector. 
Then,  the average $\mean{e^{in\phi_j}}$ vanishes by symmetry [since 
$\phi_j$ is measured in the laboratory, not with respect to the reaction 
plane, $\mean{e^{in\phi_j}}$ does not correspond to the flow $v_n$ 
defined in Eq.~(\ref{defvn})]: 
the first term on the right-hand side (r.-h.~s.) of Eq.~(\ref{defc2}) 
vanishes and the cumulant reduces to the measured two-particle 
correlation. 

The relevance of the cumulant appears when considering the more 
realistic case of a detector with uneven acceptance. 
Then, the first term on the r.-h.~s. of Eq.~(\ref{defc2}) 
can be nonvanishing. But the cumulant vanishes if 
$\phi_1$ and $\phi_2$ are uncorrelated. 
Thus the cumulant $\cumul{e^{in(\phi_1-\phi_2)}}$ 
isolates the physical correlation, and disentangles it 
from trivial detector effects. 

There are several physical contributions to the correlation
$\cumul{e^{in(\phi_1-\phi_2)}}$, which separate into flow 
and nonflow (or direct) correlations. 
When the source is isotropic (no flow), only direct correlations 
remain. 
They scale with the multiplicity $M$ like $1/M$ 
\cite{Dinh:2000mn,Borghini:2000cm},
as can be easily understood when considering correlations between 
the decay products of a resonance:
when a $\rho$ meson decays into two pions, momentum conservation 
induces an angular correlation of order unity between the decay pions;
besides, the probability that two arbitrary pions seen in the detector 
result from the same $\rho$ decay scales with the total number of 
pions like $1/M$. 
All in all, the correlation between two arbitrary pions is of order $1/M$. 
If the source is not isotropic, flow, which is by definition a correlation 
between emitted particles and the reaction plane, generates azimuthal 
correlations between any two outgoing 
particles, and gives a contribution $v_n^2$ to the second-order
cumulant, as will be explained in Sec.\ \ref{s:int_flow}. 
One can measure the flow using the second-order cumulant if this 
contribution dominates over the nonflow contribution, i.e., 
if $v_n\gg 1/\sqrt{M}$ \cite{Dinh:2000mn,Borghini:2000cm}. 
This is the domain of validity of standard flow analyses, which are 
based on two-particle correlations.

Our main point is that through the construction of higher order 
cumulants, one can separate flow and nonflow correlations. 
To illustrate how this works, 
we consider for simplicity a perfect 
detector. Then, we decompose the measured four-particle correlation  
as follows:
\begin{equation}
\label{defc4bis}
\mean{e^{in(\phi_1+\phi_2-\phi_3-\phi_4)}}  
= 
\mean{e^{in(\phi_1-\phi_3)}}\mean{e^{in(\phi_2-\phi_4)}} + 
\mean{e^{in(\phi_1-\phi_4)}}\mean{e^{in(\phi_2-\phi_3)}} +
\cumul{e^{in(\phi_1+\phi_2-\phi_3-\phi_4)}}.
\end{equation}
If the particles are correlated pairwise, there are two possible 
combinations leading to a nonvanishing value for the left-hand side 
(l.-h.~s.):
the pairs can be either (1,3) and (2,4), or (1,4) and (2,3). 
This yields the first two terms in the right-hand side. 
The remaining term 
$\cumul{e^{in(\phi_1+\phi_2-\phi_3-\phi_4)}}$, which is by 
definition the fourth-order cumulant, 
is thus insensitive to two-particle nonflow correlations. 
However, it may still be influenced by higher order nonflow 
correlations: if, for instance, a resonance decays into four 
particles, the resulting correlations between the reaction products  
do not factorize as in Eq.~(\ref{defc4bis}). 
We call such correlations ``direct'' four-particle correlations. 
Fortunately, their contribution to the fourth-order cumulant is 
very small: it scales with the multiplicity like 
$1/M^3$\cite{Borghini:2001sa}, while the measured correlation 
$\mean{e^{in(\phi_1+\phi_2-\phi_3-\phi_4)}}$ is generally much 
larger, of order $1/M^2$ 
[the two-particle correlation terms in the r.-h.~s. of 
Eq.~(\ref{defc4bis}) are of order $1/M$, as explained above]. 
On the other hand, flow yields a contribution $-v_n^4$
to the cumulant, as we shall see in Sec.\ \ref{s:int_flow}.
Therefore, the cumulant is dominated by the flow as soon as 
$v_n \gg 1/M^{3/4}$. 
This is a major improvement on two-particle correlations, which are 
limited by the much stronger constraint $v_n \gg 1/\sqrt{M}$.

Equation (\ref{defc4bis}) can be rewritten 
\begin{equation}
\label{defc4}
\mean{e^{in(\phi_1+\phi_2-\phi_3-\phi_4)}}  
=2\,\mean{e^{in(\phi_1-\phi_3)}}^2 + 
\cumul{e^{in(\phi_1+\phi_2-\phi_3-\phi_4)}},
\end{equation}
where we have used the symmetry between $\phi_1$ and $\phi_2$ 
(resp.\ $\phi_3$ and $\phi_4$). 
However, Eqs.~(\ref{defc4bis}) and (\ref{defc4}) only 
hold for a perfect detector, therefore they are of little 
practical use. 
It is in fact possible to build an expression for the fourth-order cumulant 
which eliminates both detector effects and nonflow correlations, 
but this expression is very lengthy. 
This is the reason why we introduce a generating function of cumulants in 
Sec.\ \ref{s:int_gene}. 
It will enable us to construct easily cumulants of arbitrary orders
for arbitrary detectors. 

More generally, the cumulant $\langle\!\langle
\exp[in(\phi_1+\cdots+\phi_k-\phi_{k+1}-\cdots-\phi_{k+l})]
\rangle\!\rangle$, 
which involves $k+l$ particles, is of order $M^{1-k-l}$ when there 
is no flow. It eliminates all nonflow correlations up to order $k+l-1$. 
Only direct correlations between $k+l$ particles remain. 
Cumulants with $k\not= l$ vanish for a perfect detector and 
are physically irrelevant. 
The interesting cumulants are the ``diagonal'' ones, with $k=l$, 
as in Eqs.~(\ref{defc2}) 
and (\ref{defc4}). The contribution of flow to these cumulants,
proportional to $v_n^{2k}$, will be evaluated precisely in 
Sec.\ \ref{s:int_flow}. 
When this contribution dominates over the nonflow contribution, 
the measured cumulant yields an estimate of the value of $v_n$, which 
we denote by $v_n\{2k\}$, where $k>0$ is in principle arbitrary.

\subsection{Generating function}
\label{s:int_gene}

Cumulants can be expressed elegantly, 
and without assuming a perfect detector as in Eq.~(\ref{defc4}), 
using the formalism of generating functions. 
For each event, we define the real-valued function $G_n(z)$, which 
depends on the complex variable $z=x+iy$:
\begin{eqnarray}
\label{newG0}
G_n(z) &=& \prod_{j=1}^M 
\left( 1+ {z^* e^{in\phi_j} + ze^{-in\phi_j}\over M} \right)\cr
&=& \prod_{j=1}^M 
\left( 1+ {2\,x \cos(n\phi_j) + 2\, y\sin(n\phi_j)\over M} \right),
\end{eqnarray}
where $z^*\equiv x-iy$ denotes the complex conjugate. 
This generating function can then be averaged over events with the same 
multiplicity $M$. 
We denote this statistical average by $\mean{G_n(z)}$.
Its expansion in power series generates measured azimuthal correlations 
to all orders:
\begin{eqnarray}
\label{expansionnewG0}
\mean{G_n(z)} &=&
1 + {z\over M}\mean{\sum_{j=1}^M e^{-in\phi_j}} + 
{z^*\over M}\mean{\sum_{j=1}^M e^{in\phi_j}}  \cr
& &\qquad + {z^2\over M^2}
\mean{\sum_{j< k}e^{-in(\phi_j+\phi_k)}} + 
{z^{*2}\over M^2}\mean{ \sum_{j<k} e^{in(\phi_j+\phi_k)}} +
{zz^*\over M^2}\mean{\sum_{j\not= k}e^{in(\phi_j-\phi_k)}}+ \cdots\cr
&=&1 + z\mean{e^{-in\phi_1}} + z^*\mean{e^{in\phi_1}}  
+ {M-1\over M}\left({z^2\over 2}
\mean{e^{-in(\phi_1+\phi_2)}} + 
{z^{*2}\over 2}\mean{ e^{in(\phi_1+\phi_2)}} +
zz^*\mean{e^{in(\phi_1-\phi_2)}}\right) + \cdots
\end{eqnarray}
where, the averages $\mean{e^{in\phi_1}}$, 
$\mean{ e^{in(\phi_1-\phi_2)}}$, etc. are defined as in 
Sec.\ \ref{s:int_cumul}. 
More generally, expanding $\mean{G_n(z)}$ to order $z^{*k}z^l$ 
yields, up to a numerical coefficient, the ($k+l$)-particle correlation
$\mean{\exp[in(\phi_1+\cdots+\phi_k-\phi_{k+1}-\cdots-\phi_{k+l})]}$. 
The generating function $\mean{G_n(z)}$ thus contains all the 
information on measured multiparticle azimuthal correlations. 

If the detector is perfect, the statistical average $\mean{G_n(z)}$ does not 
depend on the phase of $z$; it only depends on $|z|=\sqrt{x^2+y^2}$. 
To see this, one may note that changing $z$ into $z\, e^{in\alpha}$ in
the generating function (\ref{newG0}) amounts to shifting all angles
by the same quantity $\phi_j\to \phi_j-\alpha$.
Now, the probability for an event to occur is unchanged under a global
rotation; therefore $\mean{G_n(z)}$ is unchanged under this rotation, 
hence the result. 
In this case, the only terms which remain in the power-series 
expansion are the isotropic terms, proportional to $z^k z^{*k}$, 
which involve only relative angles. 

The generating function provides us with a way to obtain a compact 
expression for cumulants of arbitrary orders. 
We define the generating function of the cumulants ${\cal C}_n(z)$ by 
\begin{equation}
\label{defc0}
{\cal C}_n(z)\equiv
M\left(\mean{G_n(z)}^{1/M}-1\right).
\end{equation}
The expansion of this function in power series of $z$ and $z^*$
defines the cumulants:
\begin{equation}
\label{defc1}
{\cal C}_n(z)\equiv
\sum_{k,l} {z^{*k}z^{l}\over k!\,l!} 
\cumul{e^{in(\phi_1+\cdots+\phi_k-\phi_{k+1}-\cdots-\phi_{k+l})}}.
\end{equation}
One easily checks 
that if the particles are uncorrelated, all the cumulants 
vanish beyond order one, i.e., for $k+l\ge 2$. 
Indeed, if all the $\phi_j$ in Eq.~(\ref{newG0}) are independent
from each other, the mean value of the product is the product of the mean
values, so that 
\begin{equation}
\mean{G_n(z)}=
\left( 1+ {z^* \mean{e^{in\phi}} + z\mean{e^{-in\phi}}\over M} \right)^M.
\end{equation}
The generating function of cumulants,  Eq.~(\ref{defc0}), then reduces to
\begin{equation}
\label{cumulvanish}
{\cal C}_n(z)= z^* \mean{e^{in\phi}} + z\mean{e^{-in\phi}}. 
\end{equation}
Comparing with Eq.~(\ref{defc1}), cumulants of order 2 and higher 
vanish when particles are uncorrelated, as expected. 

The cumulant $\cumul{e^{in(\phi_1-\phi_2)}}$ obtained when expanding 
Eqs.~(\ref{defc0}) and (\ref{defc1}) to order $zz^*$ coincides with the 
second-order cumulant defined in Eq.~(\ref{defc2}) in the limit of large $M$ 
(see Appendix \ref{s:cumulants}). 
Expanding ${\cal C}_n(z)$ to order $z^2z^{*2}$, one obtains an 
expression for the cumulant 
$\cumul{e^{in(\phi_1+\phi_2-\phi_3-\phi_4)}}$ 
which reduces to Eq.~(\ref{defc4}) for a perfect detector. 
But the expression derived from Eqs.~(\ref{defc0}) and 
(\ref{defc1}) is still valid 
with an imperfect detector, while  Eq.~(\ref{defc4}) is not. 

As mentioned in Sec.\ \ref{s:int_cumul}, cumulants with $k\not= l$ 
vanish for a perfect detector, since the generating function 
${\cal C}_n(z)$ in Eq.~(\ref{defc1}) depends only on $|z|$. 
The interesting cumulants are the diagonal terms with $k=l$, 
which are related to the flow. We denote them by $c_n\{2k\}$:
\begin{equation}
\label{defc2k}
c_n\{2k\}\equiv \cumul{e^{in(\phi_1+\cdots+\phi_k-\phi_{k+1}-\cdots-\phi_{2k})}}.
\end{equation}
In practice, expanding the generating function ${\cal C}_n(z)$ 
analytically is rather tedious beyond order 2. 
The simplest way to extract $c_n\{2k\}$ is to tabulate the
generating function (\ref{defc0}), and then compute numerically the 
coefficients of its power-series expansion, using interpolation formulas 
which can be found in Appendix \ref{s:interpoli}. 

Finally, we have assumed here that the multiplicity $M$ is exactly the same 
for all events involved in the analysis. 
In practice, one performs the flow analysis for a class of events belonging 
to the same centrality interval, and $M$ fluctuates from one event to the 
other. 
That explains our introducing the factor $1/M$ in the definition of the 
generating function (\ref{newG0}), as explained in more detail in 
Appendix\ \ref{s:cumulants}. 
The average over events $\mean{G_n(z)}$ then involves an average over $M$, 
and $M$ must be replaced by its average value $\mean{M}$ in the definition
of the cumulants, Eq.~(\ref{defc0}). 
This, however, leads to errors, especially when the acceptance is bad (see 
Appendix\ \ref{s:cumulants}). 
This point will also be illustrated by the simulations presented in 
Sec.\ \ref{s:mc}.  
In order to avoid these effects, one may use only a randomly 
selected subset of the detected particles, with a fixed multiplicity 
$M$, to construct the generating function.

\subsection{Contribution of flow to the cumulants}
\label{s:int_flow}

Let us evaluate the contribution of flow to the cumulants $c_n\{2k\}$. 
For simplicity, we assume that the detector is perfect. 
The generalization to an uneven acceptance is performed in 
Appendix \ref{s:accint}.
Under this assumption, one easily calculates the generating function 
$\mean{G_n(z)}$ and, from it, the values of the cumulants. 
 
Let us call $\Phi_R$ the azimuthal angle of the reaction plane of a given 
event. The average over events can be performed in two steps: one
first estimates the average over all events with the same $\Phi_R$; 
then one averages over $\Phi_R$. 
We denote by $\mean{x|\Phi_R}$ the average of a quantity $x$ 
for fixed $\Phi_R$. 
With this notation, the definition of $v_n$, Eq.~(\ref{defvn}), gives
\begin{equation}
\label{vnphir}
\mean{e^{in\phi_j}|\Phi_R}=v_n\,e^{in\Phi_R}.
\end{equation}
Neglecting for simplicity nonflow correlations between particles, 
we obtain from Eq.~(\ref{newG0}) 
\begin{equation}
\label{meanG0}
\mean{G_n(z)|\Phi_R}  = 
\left( 1+ {z\, v_n\, e^{-in\Phi_R} + z^*v_n\, e^{in\Phi_R}\over M} \right)^M .
\end{equation}
One must then average over $\Phi_R$: 
\begin{equation}
\label{avphiR}
\mean{G_n(z)} 
=\int_0^{2\pi}\mean{G_n(z)|\Phi_R}  
{d\Phi_R\over 2\pi}.
\end{equation}
Inserting Eq.~(\ref{meanG0}) in this expression, one obtains 
\begin{eqnarray}
\label{meanG01}
\mean{G_n(z)} &=&
\sum_{k=0}^{[M/2]} {M!\over(M-2k)! (k!)^2 } \left({v_n\over
M}\right)^{2k} |z|^{2k}\cr
&\simeq& I_0(2\,v_n\,|z|), 
\end{eqnarray}
where, in the last equation, we have assumed that $M$ is large, so 
that $M!/(M-2k)!=M^{2k}$ and one may extend the sum over $k$ to 
infinity. $I_0$ denotes the modified 
Bessel function of the first kind. 
The result depends only on $|z|$, as expected from the
discussion in Sec.\ \ref{s:int_flow}. 

The generating function of the cumulants (\ref{defc0}) now reads 
\begin{equation}
\label{cflow}
{\cal C}_n(z)\simeq M\left(I_0(2v_n|z|)^{1/M}-1\right)\simeq
\ln I_0(2\, v_n\, |z|).
\end{equation} 
This equation can be expanded in power series. Comparing with 
Eq.~(\ref{defc1}), the cumulants with $k\not= l$ vanish, as 
expected for a perfect detector, while the diagonal cumulants
$c_n\{ 2k\}$ defined by Eq.~(\ref{defc2k}) are related to $v_n$. 
From the measured $c_n\{ 2k\}$, one thus obtains an estimate 
of $v_n$, which is denoted by $v_n\{ 2k\}$. 
The lowest order estimates are 
\begin{mathletters}
\label{c246}
\begin{eqnarray}
v_n\{2\}^2 &\equiv& c_n\{2\},\label{c2} \\
v_n\{4\}^4 &\equiv& -c_n\{4\},\label{c4}\\
v_n\{6\}^6 &\equiv& c_n\{6\}/4.\label{c6}
\end{eqnarray}
\end{mathletters}
When the detector acceptance is far from isotropic, as is the case of the 
PHENIX detector at RHIC~\cite{Lacey:2001}, which covers
approximately half the range in azimuth, these relations no longer hold. 
The issue of acceptance corrections, discussed in detail in 
Appendix \ref{s:acceptance}, is more subtle than might be thought at 
first sight, for the following reason: 
when there is flow, the probability that a particle be detected 
depends on the orientation of the reaction plane $\Phi_R$ if 
the detector only has partial azimuthal coverage.
Hence, if a fixed number of particles are emitted, the number 
of particles seen in the detector depends of $\Phi_R$. 
Reciprocally, for a fixed value of the multiplicity $M$ seen 
in the detector, the probability distribution of $\Phi_R$ is 
not uniform, which creates an important bias in the flow analysis. 
In the calculations of Appendix \ref{s:accint}, we assume the 
centrality selection is done by an {\em independent} detector (for 
instance a zero-degree calorimeter) which has (at least approximately) 
full azimuthal coverage, so that the distribution of $\Phi_R$ is 
uniform for the sample of events used in the flow analysis. 

Under this assumption, one can derive general relations between the 
cumulants and the flow. 
It turns out that, in general,  $c_n\{2k\}$ depends not only on $v_n$, 
but also on other harmonics $v_p$ with $p\not= n$. 
In order to obtain the corresponding relations, 
we introduce the acceptance function $A(\phi)$, 
which is the probability that a particle with azimuthal angle $\phi$ 
be detected. 
The Fourier coefficients of this acceptance function are 
\begin{equation}
\label{defap}
a_p\equiv \int_{0}^{2\pi} e^{-ip\phi}A(\phi){d\phi\over 2\pi}.
\end{equation}
The relations between the cumulants $c_n\{2k\}$ and the estimates 
$v_p\{2k\}$ involve these coefficients. 
They are derived in Appendix \ref{s:accint}, to leading order in $v_p$. 
The results for directed flow and elliptic flow are given by 
Eqs.~(\ref{c24accv1})  and (\ref{c24accv2}), respectively.

\subsection{Errors}
\label{s:int_stat}

We now examine the orders of magnitude of systematic errors, arising 
from unknown nonflow correlations, and statistical errors, due to the 
finite number of events available. 
More precisely, we estimate the difference $\delta v_n\{2k\}$ 
between the true integrated flow $v_n$ and its values reconstructed 
from the cumulants, $v_n\{2k\}$, defined in Eqs.~(\ref{c246}). 
We show which value of $2k$ minimizes the total uncertainty. 

As explained in Sec.~\ref{s:int_cumul}, nonflow $2k$-particle 
correlations give a contribution of order $M^{1-2k}$ to the
cumulant $c_n\{2k\}$. 
This is to be compared with the contribution of flow derived
in Sec.\ \ref{s:int_flow}, of order $v_n^{2k}$. 
We may thus write 
\begin{equation}
\label{errsyst}
v_n\{2k\}^{2k}-v_n^{2k}={\cal O}(M^{1-2k}),
\end{equation} 
which is an estimate of the systematic error $\delta v_n\{2k\}$ due to 
nonflow correlations. 
Obviously, flow can be measured only if $v_n^{2k}\gg M^{1-2k}$.
For large orders $k\gg 1$, this condition becomes 
\begin{equation}
\label{flowlimit}
v_n\gg 1/M,
\end{equation}
which is a necessary condition for the flow to be measurable
\cite{Borghini:2001sa}. 
We believe there is no way to extract a flow of order $1/M$ or smaller. 

In this paper, we always assume that condition (\ref{flowlimit}) is 
fulfilled. If this is the case, the systematic error on $v_n$ given by 
Eq.~(\ref{errsyst}), $\delta v_n\{2k\}\sim (Mv_n)^{1-2k}$, 
becomes smaller and smaller as $k$ increases: thus 
one should construct cumulants of orders as high as possible. 

One must also take into account the statistical error, due to the 
finite number of events $\N$ available. 
The order of magnitude of statistical errors can easily be understood. 
The cumulant $c_n\{2k\}$ involves correlations between $2k$ particles 
belonging to the same event. 
There are roughly $M^{2k}$ ways (for large enough $M$) to choose $2k$ 
particles among the $M$ particles detected, and one averages over all 
possible combinations. 
Since this is done for all $\N$ events, there is a total of $M^{2k}\N$ 
subsets of $2k$ particles involved in the evaluation of the cumulants. 
The resulting statistical error is therefore 
\begin{equation}
\label{errstat}
v_n\{2k\}^{2k}-v_n^{2k}={\cal O}\left({1\over\sqrt{M^{2k}\N}}\right). 
\end{equation}
Unlike the systematic error, the statistical error 
generally increases with increasing cumulant order $2k$ 
(it may in fact decrease in some cases, but only slightly, 
see Appendix \ref{s:stat_int}). 
Therefore, the order $2k$ which gives the best compromise 
is that for which statistical and systematic errors are both of 
the same magnitude. Equating the right-hand sides of 
Eqs.~(\ref{errsyst}) and (\ref{errstat}), one obtains the
optimal cumulant order~\cite{Borghini:2001sa}:
\begin{equation}
2\,k_{\rm opt}\simeq  2+\frac{\ln \N}{\ln M}.
\label{err-stat}
\end{equation}
In most practical cases, the fourth-order cumulant ($2k=4$), that is, 
removing two-particle nonflow correlations, is to be preferred. 

Statistical errors are discussed more thoroughly in Appendix 
\ref{s:stat_int}. 
There, we derive exact formulas for the standard 
deviations of the cumulants, and for their mutual correlations. 
Two regimes can be distinguished, depending on the value of the 
dimensionless parameter $\chi\equiv v_n\sqrt{M}$, 
which has been used previously as a measure of the 
reaction plane resolution~\cite{Ollitrault:1997di}.
If $\chi\ll 1$, the standard errors agree with the simple estimate 
(\ref{errstat}), and different estimates $v_n\{2k\}$ and $v_n\{2l\}$ 
with $k\not= l$ 
are uncorrelated. If $\chi\gg 1$, on the other hand, 
they are strongly correlated and the 
standard error becomes
\begin{equation}
\label{errstatg}
\delta (v_n\{2k\})_{\rm stat}={1\over\sqrt{2M\N}},
\end{equation}
independent of the order $2k$, and much larger than the estimate 
(\ref{errstat}). 

We also discuss in Appendix \ref{s:stat_int} the non-Gaussian 
character of the fluctuations of the estimates $v_n\{2k\}$, 
due to their non-linear relations with the cumulants, 
Eqs.~(\ref{c246}). 
In particular, statistical fluctuations may result in a ``wrong''
sign for the cumulant, in the sense that $v_n\{2k\}^{2k}$ defined 
by Eqs.~(\ref{c246}) is negative. If this happens, the flow 
clearly cannot be estimated from the corresponding cumulant.

\subsection{Non-unit weights}
\label{s:weights}

In the definition of the generating function, Eq.~(\ref{newG0}), each 
particle was given the same weight. 
A more general form is 
\begin{equation}
\label{G0weights}
G_n(z) = \prod_{j=1}^M 
\left[ 1+ {w_j\over M}\left(z^* e^{in\phi_j} + ze^{-in\phi_j}\right) \right].
\end{equation}
The weight $w_j$ can be any arbitrary function of the rapidity $y$ of the 
particle, its transverse momentum $p_T$, its mass. 

In order to obtain the highest accuracy on the flow measurement, 
$w_j$ should be chosen proportional to the flow itself, as shown in 
\cite{Borghini:2001sa} (see also \cite{Tsang:1991,Danielewicz:1994nb}):
the ideal weight is  $w(p_T,y)\propto v'_n(p_T,y)$,
which is intuitively clear: one must give higher 
weights to particles with stronger flow. 
This is also the best choice if one uses the standard method, 
involving the determination of the reaction plane. 

Ideally, the flow analysis should be performed twice: a first 
measurement of the flow can be done with reasonable guesses for the 
weights; measuring differential flow as a function of $y$ and $p_T$ 
for various particles (see Sec.~\ref{s:differential}), the values 
obtained can be used as the new weights in a second, more accurate 
analysis. This is the procedure recently followed in \cite{Chung:2000ny}.

A variety of weights have been used in analyses of the directed flow $v_1$. 
Since $v_1$ changes sign at midrapidity, the weight must be an odd function 
of the rapidity in the center-of-mass frame. 
Most often, the weight is simply the sign of $y$, 
with~\cite{Danielewicz:1985hn} or without~\cite{Doss:1986eh}
a gap at midrapidity. 
A linear dependence in $y$ was used 
in~\cite{Gosset:1989cm,Ogilvie:1989a,Sullivan:1990}. 
The latter choice is better, since $v_1$ is itself linear near 
midrapidity. 
The transverse momentum dependence of $w$ is most often linear, as in the 
original paper~\cite{Danielewicz:1985hn}. 
Unit weights, independent of $p_T$, are also widely 
used~\cite{Taranenko:1999yh,Appelshauser:1998dg,Fai:1987bt,Htun:1999nc}.
They are convenient, because no particle identification is required. 
However, since $v_1$ is linear in $p_T$, (at least at low $p_T$ 
\cite{Danielewicz:1994nb}), the original choice $w\propto p_T$ is likely 
to give more accurate results, although the opposite conclusion was reached 
in~\cite{Fai:1987bt}. 
At intermediate energies, one can in addition choose a weight proportional
to the mass of the particle, to take into account the fact that 
nuclear fragments flow more than protons~\cite{Ogilvie:1989b,Krofcheck:1989}.

Unlike directed flow, elliptic flow is an even function of the center-of-mass 
rapidity: therefore the weights are usually chosen independent of rapidity. 
The weights are either independent of transverse momentum 
\cite{Ackermann:2001tr,Poskanzer:1998yz,Appelshauser:1998dg,Aggarwal:1997iu,Raniwala:2000yz}
or proportional to $p_T^2$
\cite{Ollitrault:1992bk,Tsang:1991,Wilson:1992}. The latter choice 
is more appropriate if $p_T$ is measured, since $v_2$ is proportional 
to $p_T^2$ at low $p_T$ \cite{Danielewicz:1994nb}.
At ultrarelativistic energies, however, $v_2$ is almost linear 
in $p_T$ above 100~MeV \cite{Ackermann:2001tr,Huovinen:2001cy}. 
A better choice may be for instance $w(p_T)=\sqrt{p_T^2+p_0^2}-p_0$, 
with $p_0\simeq 100$~MeV: this weight, quadratic at low $p_T$ and 
linear at high $p_T$, reproduces roughly the measured $p_T$ 
dependence of $v_2$. 

With a non-unit weight $w$, the following modifications 
should be applied:
\begin{itemize}
\item{In the relations between the cumulants and the flow, 
i.e., Eqs.~(\ref{c246}) for integrated flow, Eqs.~(\ref{m=1})
and (\ref{m=2}) for 
differential flow, and the equations of Appendix \ref{s:acceptance}, 
$v_n$ now stands for $\mean{w e^{in(\phi-\Phi_R)}}$. 
Comparing with the previous definition, Eq.~(\ref{defvn}), 
the flow is now weighted by $w$, as can be expected from 
the definition of the generating function, Eq.~(\ref{G0weights}).}

\item{In the formulas giving the statistical errors, 
derived in Appendix \ref{s:statistical}, one must in addition replace 
$M$ by $M/\mean{w^2}$. 
As a consequence, the resolution $\chi^2$ appearing in 
Eqs.~(\ref{erstatint}), (\ref{corint}), (\ref{statm=1}) and 
(\ref{statm=2}) 
now stands for $\chi^2=M\mean{wv_n}^2/\mean{w^2}$. }

\item{In the interpolation formulas of Appendix \ref{s:interpolation}, 
the value of $r_0$ [Eq.~(\ref{defxy})] should be scaled by the 
typical value of $w$, for instance by $\sqrt{\mean{w^2}}$.}
\end{itemize}

\section{Differential flow}
\label{s:differential}

When one has measured the flow integrated over phase space, the next step
 is to move on to the ``differential flow'' analysis, i.e., the measurement 
of flow in a narrower phase space window. 
We call a particle belonging to the window of interest a ``proton'' (although 
it can be anything else). 
We denote by $\psi$ its azimuthal angle, and $v'_n$ its flow harmonics 
(the so-called ``differential flow''), 
$v'_n\equiv\mean{e^{in(\psi-\Phi_R)}}$. 
The particles used to estimate the integrated flow $v_n$ are named ``pions''.  

In order to measure the differential flow of the protons, we correlate their 
azimuth $\psi$ with the pion azimuths $\phi_j$. 
Once the integrated flow $v_n$ is known, one can reconstruct the differential 
flow $v'_n$ from this correlation, 
and also higher harmonics $v'_{2n}$, $v'_{3n}$, etc. It follows 
that differential elliptic flow, $v'_2$, can be reconstructed using 
either integrated directed flow or integrated elliptic flow. 
Following  \cite{Poskanzer:1998yz}, we denote by $v'_{p/n}$ the 
differential flow $v'_p$ measured with respect to integrated flow 
$v_n$, where $p$ is a multiple of $n$. 
At intermediate energies, one usually measures $v'_{2/1}$ 
\cite{Taranenko:1999yh} while $v'_{2/2}$ is more accurate 
at ultrarelativistic energies where $v_1$ becomes very 
small~\cite{Appelshauser:1998dg}. 

The differential flow is reconstructed by taking the cumulants 
of azimuthal correlations between the proton and the pions. 
These are constructed in Sec.\ \ref{s:diff_cumul} by means 
of an appropriate generating function. 
The subtraction of ``autocorrelations'', in the case when 
the proton is one of the pions, is briefly discussed 
in Sec.\ \ref{s:diff_auto}.
In Sec.\ \ref{s:diff_flow}, we derive the relations between the 
cumulants and the differential flow, $v'_{p/n}$. As in the 
case of integrated flow, cumulants of different orders yield 
different estimates of $v'_{p/n}$. 
The optimal choice is that which minimizes uncertainties,
discussed in Sec.\ \ref{s:diff_stat}. 

\subsection{Cumulants}
\label{s:diff_cumul}

To measure a proton differential flow $v'_p$, we first construct a 
generating function of measured azimuthal correlations between the 
proton and pions. 
This function is the average value over all protons of $e^{ip\psi}G_n(z)$, 
where $G_n(z)$ is the generating function defined by Eq.~(\ref{newG0}), 
evaluated for the event where the proton belongs. 
Note that the average procedure is not exactly the same as when studying
integrated flow. One must first average over the protons in a same event 
[i.e., with the same $G_n(z)$]; then, average over the events where there 
are ``protons'' (if the ``proton'' is some rare particle, 
or if the phase space window is small, there may be
events without a proton).

Expanding in power series of $z$ and $z^*$, one obtains 
\begin{equation}
\label{expansionnewGm}
\mean{e^{ip\psi}G_n(z)} =
\mean{e^{ip\psi}} + z\mean{e^{i(p\psi-n\phi_1)}} +
z^*\mean{e^{i(p\psi+n\phi_1)}}+\cdots
\end{equation}
This generates measured azimuthal correlations between the proton and 
an arbitrary number of pions. 
The generating function of the cumulants is a complex-valued function 
of the complex variable $z$:
\begin{equation}
\label{defcm}
{\cal D}_{p/n}(z)
\equiv\frac{\mean{e^{ip\psi}G_n(z)}}{\mean{G_n(z)}},
\end{equation}
where $\mean{G(z)}$ denotes an average over {\it all\/} events, 
as in Sec.\ \ref{s:int_gene}. 
The cumulants are by definition the coefficients in the 
power-series expansion of this function:
\begin{equation}
\label{defcm1}
{\cal D}_{p/n}(z)
\equiv \sum_{k,l} {z^{*k}z^{l}\over k!\,l!} 
\cumul{e^{ip\psi+in(\phi_1+\cdots+\phi_k-\phi_{k+1}-\cdots-\phi_{k+l})}}.
\end{equation}
The physical significance of these cumulants is the same as for the cumulants
used in the analysis of integrated flow. 
They eliminate detector effects and lower-order correlations, so that 
only the direct correlation between $k+l$ pions and the proton, 
of order $M^{-k-l}$, and the correlation due to flow remain. 
If the proton is not correlated with the pions, for instance, 
Eq.~(\ref{defcm}) gives 
${\cal D}_{p/n}(z)=\mean{e^{ip\psi}}$, independent of $z$, and 
all cumulants with $k+l\ge 1$ vanish according to Eq.~(\ref{defcm1}). 
In the more general case when there are correlations, 
expanding Eq.~(\ref{defcm}) to order $z$ and identifying 
with Eq.~(\ref{defcm1}), one obtains 
\begin{equation}
\label{d_p/n}
\cumul{e^{i(p\psi-n\phi_1)}}\equiv \mean{e^{i(p\psi-n\phi_1)}}-
\mean{e^{ip\psi}}\mean{e^{-in\phi_1}}. 
\end{equation}
This cumulant is analogous to Eq.~(\ref{defc2}), and can be
interpreted in the same way. 

If the detectors used to measure protons and pions are perfect, 
simplifications occur: first, the cumulants defined in 
Eq.~(\ref{defcm1}) are real. 
To show this, we use the property that if the detector is perfect, the 
probability that an event occur is unchanged when one reverses the sign 
of all azimuthal angles (i.e., $\psi\to -\psi$ and $\phi_j\to -\phi_j$), 
therefore ${\cal D}_{p/n}(z)$ is unchanged under this transformation. 
Now, the transformation $\phi_j\to -\phi_j$ amounts to changing $z$ 
into $z^*$ in $G_n(z)$, according to Eq.~(\ref{newG0}). 
Thus, Eq.~(\ref{defcm}) can also be written 
${\cal D}_{p/n}(z)=\mean{e^{-ip\psi}G_n(z^*)}/\mean{G_n(z^*)}$. 
Comparing with the original definition, Eq.~(\ref{defcm}), 
$z$ has been changed to $z^*$ and $\psi$ to $-\psi$. 
Now, changing $\psi$ to $-\psi$ in Eq.~(\ref{defcm}) 
amounts to taking the complex conjugate of ${\cal D}_{p/n}(z)$, 
since $G_n(z)$ is real. 
One finally obtains ${\cal D}_{p/n}(z)={\cal D}^*_{p/n}(z^*)$, from 
which one easily deduces that the coefficients in the expansion 
(\ref{defcm1}) are real. 
They are in general complex with a realistic detector, but only the 
real part is relevant. 

Furthermore, most cumulants vanish if the detector is perfect. 
In order to see this property, we shift  all angles by the same quantity
$\alpha$, which does not change the probability of the event. 
The angles of the pions are changed to $\phi_j-\alpha$, which amounts 
to changing $z$ into $ze^{in\alpha}$ in $G_n(z)$, as explained in 
Sec.\ \ref{s:int_gene}. 
Similarly, the angle of the proton $\psi$ is changed to
$\psi\to\psi-\alpha$, so that $e^{ip\psi}G_n(z)$ becomes 
$e^{ip\psi}e^{-ip\alpha}G_n(ze^{in\alpha})$. Averaging over $\alpha$
gives 0 unless $p$ is a multiple of $n$, which is the case we 
are interested in. Writing $p= m n$, 
the only terms which remain in the power series expansion 
of $\mean{e^{ip\psi}G_n(z)}$ are the terms in $z^{*k}z^{k+m}$. 
To obtain the generating function of cumulants ${\cal D}_{p/n}(z)$, 
one must divide by $\mean{G_n(z)}$. Since this quantity 
depends only on $|z|$, as explained in Sec.\ \ref{s:int_gene},
the only nonvanishing cumulants in Eq.~(\ref{defcm1}) 
are those with $l=k+m$. 
Other cumulants are physically irrelevant. 

Finally, the relevant quantities are 
\begin{equation}
\label{defdk}
d_{mn/n}\{2k+m+1\}\equiv
{\Re}\left[\cumul{e^{in(m\psi+\phi_{1}+...+\phi_{k}-\phi_{k+1}-...-\phi_{2k+m})}}\right].
\end{equation}
where ${\Re}$ denotes the real part, and 
the notation $\{2k+m+1\}$ means that the cumulant involves 
correlations between $2k+m+1$ particles (one proton and $2k+m$ pions).

When there is no flow, $d_{mn/n}\{2k+m+1\}$ is of order $M^{-2k-m}$. 
Flow gives a contribution to this cumulant, proportional 
to $v'_p v_n^{2k+m}$, which is calculated in Sec.\ \ref{s:diff_flow}. 
If this is the dominant contribution, one obtains an estimate of the 
differential flow $v'_p$ from the cumulant $d_{p/n}\{2k+m+1\}$, using a 
previously determined value of the integrated flow $v_n$. This estimate 
will be denoted by $v'_{p/n}\{2k+m+1\}$.

Analytical expressions of higher order cumulants, deduced from 
Eq.~(\ref{defcm1}), are lengthy. 
As in the case of integrated flow, the simplest way to extract them 
consists in tabulating the generating function ${\cal D}_{mn/n}(z)$, 
and then computing numerically the coefficients of its power-series 
expansion, through interpolation formulas which are given in 
Appendix \ref{s:accdiff}.

\subsection{Autocorrelations}
\label{s:diff_auto}

When studying the azimuthal correlations between the ``proton'' and
the ``pions'', the ``proton'' must not be one of the ``pions'', otherwise 
trivial autocorrelations would appear. 
This problem is well known in the standard flow analysis
\cite{Danielewicz:1985hn}: in order to avoid it, one excludes 
the particle under study (the ``proton'') from the definition 
of the flow vector used to estimate the reaction plane, which 
usually involves all the other particles (the ``pions''). 
Here,
if the proton is one of the pions, one simply 
removes its  contribution by dividing $G_n(z)$ by 
$1+(z^*e^{in\psi}+z e^{-in\psi})/M$ in the numerator of 
Eq.~(\ref{defcm}). 
The generalization of this procedure to non-unit weights is 
straightforward.

\subsection{Contribution of flow to the cumulants}
\label{s:diff_flow}

Let us now calculate the contribution of flow to the cumulants
$d_{mn/n}\{2k+m+1\}$. 
As in Sec.\ \ref{s:int_flow}, we neglect nonflow correlations 
and assume a perfect detector for simplicity.
Under these assumptions, we can compute the generating 
function of the cumulants ${\cal D}_{p/n}(z)$. 
We first average over all events with a fixed orientation of the reaction 
plane $\Phi_R$, and over the protons in each single event:
\begin{equation}
\mean{e^{ip\psi}|\Phi_R}=v'_p e^{ip\Phi_R}. 
\end{equation}
Hence, Eq.~(\ref{defcm}) becomes 
\begin{equation}
\label{cpflow}
{\cal D}_{p/n}(z) = 
\frac{\int_0^{2\pi}e^{ip\Phi_R}\mean{G_n(z)|\Phi_R}d\Phi_R/2\pi}
{\mean{G_n(z)}}\, v'_p, 
\end{equation}
where the denominator is given by Eq.~(\ref{meanG01}), and 
$\mean{G_n(z)|\Phi_R}$ by Eq.~(\ref{meanG0}). 
The numerator vanishes unless $p$ is a multiple of $n$, i.e., 
$p=mn$ with $m$ integer. 
Integrating over $\Phi_R$, one then obtains:
\begin{eqnarray}
\label{meanGm}
\int_0^{2\pi}e^{imn\Phi_R}\mean{G_n(z)|\Phi_R}{d\Phi_R\over 2\pi}&=&
\sum_{k=0}^{[(M+m)/2]} {M!\over(M-m-2k)! \,k!\, (2k+m)! } 
\left({v_n\over M}\right)^{2k+m} z^{*k}z^{k+m}\cr
&\simeq& I_m(2\,v_n\,|z|)\left(\frac{z}{|z|}\right)^m,
\end{eqnarray}
where, in the last identity, we have assumed that $M$ is large, so that 
$M!/(M-m-2k)!\simeq M^{2k+m}$ and we may extend the sum over $k$ 
to infinity. 
Equation (\ref{cpflow}) gives
\begin{equation}
\label{genfuncdiffflow}
{\cal D}_{mn/n}(z) = \frac{I_m(2|z|v_n)} {I_0(2|z|v_n)} 
\, \left(\frac{z}{|z|}\right)^m v'_{mn}.
\end{equation}
This equation can be expanded in power series of $z$ and $z^*$. 
The coefficients of the power-series expansion are real, as 
expected from the discussion in Sec.\ \ref{s:diff_cumul}. 
Comparing with Eq.~(\ref{defcm1}), cumulants with $l\not= k+m$ 
vanish, which was also expected; 
cumulants with $l=k+m$, which are the $d_{mn/n}\{2k+m+1\}$ introduced 
in Eq.~(\ref{defdk}), are proportional to $v'_{mn}$. We thus 
obtain estimates of the differential flow $v'_{mn}$, which we denote 
by $v'_{mn/n}\{2k+m+1\}$, from the measured cumulants.  
For $m=1$ (the proton is correlated with pions in the same flow
harmonic) these estimates are given by 
\begin{mathletters}
\label{m=1}
\begin{eqnarray}
v'_{n/n}\{2\}&\equiv & d_{n/n}\{2\}/v_n\label{d_n/n2}\\
v'_{n/n}\{4\}&\equiv & -d_{n/n}\{4\}/v_n^3,\label{d_n/n4}
\end{eqnarray}
\end{mathletters}
while for $m=2$ (useful when measuring differential elliptic flow $v'_2$ 
with respect to the integrated directed flow $v_1$)
\begin{mathletters}
\label{m=2}
\begin{eqnarray}
v'_{2n/n}\{3\}&\equiv &  d_{2n/n}\{3\}/v_n^2\label{d_2n/n2}\\
v'_{2n/n}\{5\}&\equiv & -d_{2n/n}\{5\}/(2\,v_n^4).\label{d_2n/n4}
\end{eqnarray}
\end{mathletters}
As in the case of integrated flow, a non-perfect acceptance may induce 
interference between the various harmonics $v'_n$, $v'_{2n}$, etc., 
modifying Eqs.~(\ref{m=1}) and (\ref{m=2}). 
The corresponding relations between the cumulants and the flow are derived 
in Appendix \ref{s:accdiff}. 
We have taken into account the possibility that integrated and 
differential flows may be measured using detectors with different 
azimuthal coverages, which is often the case in practice (see for 
instance \cite{Taranenko:1999yh}).
In particular, if the detector used for integrated flow is perfect, 
we show that no correction is required for $v'_n$, whatever the 
detector used for differential flow may be, which is intuitively 
obvious: the only difference when using a smaller detector for 
the reconstruction of differential flow is then a loss in ``proton'' 
multiplicity, resulting in higher statistical errors, which we now 
discuss.

\subsection{Errors}
\label{s:diff_stat}

As in Sec.\ \ref{s:int_stat}, we now evaluate the contributions of 
nonflow correlations and statistical fluctuations to the cumulants. 
This will allow us to determine the optimal cumulant order to be used, 
which minimizes the total uncertainty on $v'_p$. 

As discussed in Sec.~\ref{s:diff_cumul}, nonflow correlations 
give an unknown contribution of order $M^{-2k-m}$ to the cumulant 
$d_{p/n}\{2k+m+1\}$, which must be compared with the contribution 
of flow, proportional to $v'_{mn}\{2k+m+1\}\, v_n^{2k+m}$ as shown 
in Sec.~\ref{s:diff_flow}. 
The systematic error on $v'_{mn/n}$ thus reads
\begin{equation}
\label{errsystd}
v'_{mn/n}\{2k+m+1\}-v'_{mn/n}={\cal O}\left((Mv_n)^{-2k-m}\right).
\end{equation}
According to Eq.~(\ref{flowlimit}), 
this systematic error becomes smaller and smaller 
as $k$ increases: the same behavior was observed for 
the systematic error on the integrated flow in 
Sec.\ \ref{s:int_stat}.

The order of magnitude of statistical errors can be estimated 
in the same way as for integrated flow. 
The cumulant $d_{p/n}\{2k+m+1\}$ involves correlations of the proton with 
$2k+m$ pions belonging to the same event. 
There are roughly $M^{2k+m}$ ways (if $M$ is large enough) to choose 
$2k+m$ pions among $M$.
Denoting by $\Np$ the total number of protons involved 
in the analysis, there is a total of $M^{2k+m}\Np$ 
subsets of particles involved in evaluating the cumulants. 
The resulting statistical error is therefore 
\begin{equation}
\label{errstatd}
v'_{mn/n}\{2k+m+1\}-v'_{mn/n}={\cal O}\left(
 {1\over\sqrt{M^{2k+m}\Np}}\right){1\over v_n^{2k+m}}.
\end{equation}
As was the case for integrated flow, the statistical error usually increases 
with the order of the cumulant, while the systematic error decreases. 
Therefore, the cumulant order which gives the best compromise 
is that for which statistical and systematic errors are both of 
the same order of magnitude. Equating the right-hand sides of 
Eqs.~(\ref{errsystd}) and (\ref{errstatd}), one obtains the
optimal cumulant order~\cite{Borghini:2001sa}:
\begin{equation}
(2\,k+m+1)_{\rm opt}\simeq  1+\frac{\ln \Np}{\ln M}.
\label{err-statd}
\end{equation}

Statistical errors are evaluated in detail in Appendix \ref{s:stat_diff}. 
We show that the simple estimate (\ref{errstatd}) is correct only if 
$\chi\equiv v_n\sqrt{M}\ll 1$. 
In this limit, different estimates $v'_{mn/n}\{2k+m+1\}$  
and $v'_{mn/n}\{2l+m+1\}$ with $k\not= l$ are uncorrelated. 
When $\chi\gg 1$, the correlation becomes strong and 
the standard error is approximately 
\begin{equation}
\label{errstatgd}
\delta (v'_{mn/n}\{2k+m+1\})_{\rm stat}={1\over\sqrt{2\Np}},
\end{equation}
independent of the order $2k+m+1$, and much larger than the estimate 
(\ref{errstatd}).

\section{Comparison with other methods}
\label{s:previousmethod}

The method proposed in this paper is closely related to our first 
cumulant-based method \cite{Borghini:2001sa}, since it is aimed at 
correcting some of the latter's limitations.
This is discussed in Sec.\ \ref{s:ourmethod}. 
Then, in Sec.\ \ref{s:standardmethods}, we compare our method with 
the two-particle correlation technique \cite{Wang:1991qh} and with 
the subevent method \cite{Danielewicz:1985hn,Poskanzer:1998yz}. 

\subsection{Comparison with our previous method}
\label{s:ourmethod}

The cumulant expansion proposed  in \cite{Borghini:2001sa} was based 
on the flow vector rather than on particles themselves. 
The flow vector is defined for each event by 
\cite{Danielewicz:1985hn,Poskanzer:1998yz} 
\begin{equation}
\label{flowvector}
Q_n\equiv\frac{1}{\sqrt{M}}\sum_{j=1}^M e^{in\phi_j}.
\end{equation}
A generating function was then defined:
\begin{equation}
\label{oldG0}
{\cal G}_0(z)\equiv \mean{e^{zQ_n^*+z^*Q_n}}.
\end{equation}
The expansion of this generating function generates all the 
moments of the distribution of $Q$, that is, $\langle Q^kQ^{*l}\rangle$. 
One easily shows that ${\cal G}_0(z)$  is closely 
related to the generating function used in the present paper. 
Using the identity
\begin{equation}
1+{z^* e^{in\phi_j}+z e^{-in\phi_j}\over M}\simeq 
\exp\left({z^* e^{in\phi_j}+z e^{-in\phi_j}\over M}\right),
\end{equation}
valid for large $M$, we may rewrite the generating function 
(\ref{newG0}) as 
\begin{eqnarray}
\label{newweight}
G_n(z)&\simeq&
\exp\left(z^* \sum_{j=1}^Me^{in\phi_j}+z\sum_{j=1}^Me^{-in\phi_j}\right)\cr
&=&\exp\left({z^*\over\sqrt{M}} Q_n+{z\over\sqrt{M}}Q_n^*\right). 
\end{eqnarray}
Thus the average over events $\mean{G_n(z)}$ coincides with 
${\cal G}_0(z/\sqrt{M})$. 
This shows that both methods are equivalent in the large $M$ limit, 
up to a rescaling of the variable $z$. 

The difference between the two approaches can be easily understood 
by expanding the two generating functions in powers 
of $z$ and $z^*$. To second order, for instance, 
Eqs.~(\ref{flowvector}) and (\ref{oldG0}) give 
\begin{equation}
\label{autocorr}
{\cal G}_0(z)= \cdots 
+{zz^*\over M}\mean{\sum_{j,k}e^{in(\phi_j-\phi_k)}}
+{z^2\over 2M}\mean{\sum_{j,k}e^{-in(\phi_j+\phi_k)}}
+{z^{*2}\over 2M}\mean{\sum_{j,k}e^{in(\phi_j+\phi_k)}}
+\cdots 
\end{equation}
while our new generating function gives (see Eq.(\ref{expansionnewG0}))
\begin{equation}
\mean{G_n(z)}= \cdots 
+{zz^*\over M^2}\mean{\sum_{j\not= k}e^{in(\phi_j-\phi_k)}}
+{z^2\over 2M^2}\mean{\sum_{j\not= k}e^{-in(\phi_j+\phi_k)}}
+{z^{*2}\over 2M^2}\mean{\sum_{j\not= k}e^{in(\phi_j+\phi_k)}}
+\cdots 
\end{equation}
The essential difference is the restriction $j\not= k$ in the sums: 
our new method is free from the autocorrelations corresponding to the 
terms with $j=k$ in Eq.~(\ref{autocorr}). 
This remains true to higher orders in $z$ and $z^*$. 

Autocorrelations have two effects:
in the first term of Eq.~(\ref{autocorr}), they give a constant, 
trivial contribution which must then be removed. 
This fixes the choice of the weight $1/\sqrt{M}$ in the definition 
of the flow vector, Eq.~(\ref{flowvector}). With another weight, 
the contribution of autocorrelations would depend on $M$, and it 
would not be easy to subtract them when $M$ is allowed to fluctuate.
With the new generating function, we are free to choose another weight, 
and we show in Appendix \ref{s:cumulants} that the weight $1/M$ 
[Eq.~(\ref{newG0})] gives more accurate results when $M$ is allowed 
to fluctuate.

In the second and third terms of Eq.~(\ref{autocorr}), autocorrelations 
create terms in $e^{\pm 2in\phi_j}$ which interfere with higher 
harmonics. 
This was the main limitation of the method exposed 
in \cite{Borghini:2001sa}. 
Eliminating all autocorrelations represents a major improvement. 
In particular, our method should enable to measure directed flow 
at RHIC, if any.

Finally, let us comment on our definition of the cumulants through 
the generating function ${\cal C}_n(z)$ in Eq.~(\ref{defc0}). 
This definition ensures that cumulants of order 2 and higher 
vanish if particles are uncorrelated [see the discussion following 
Eq.~(\ref{defc1})]. 
In the limit when $M$ is large, one recovers 
${\cal C}_n(z)=\ln\mean {G_n(z)}$, in agreement with the 
standard definition of cumulants in probability theory
\cite{vanKampen}, and with the definition we adopted in 
\cite{Borghini:2001sa}.

\subsection{Comparison with standard methods}
\label{s:standardmethods}

In \cite{Wang:1991qh}, it was proposed to analyse flow 
using two-particle azimuthal correlations. 
More specifically, one defines by $P_{\rm corr}(\Delta\phi)$
[resp. $P_{\rm uncor}(\Delta\phi)$] the distribution 
of the relative angle $\Delta\phi\equiv \phi_1-\phi_2$, 
where $\phi_1$ and $\phi_2$ are any two particles belonging 
to the same event (resp. belonging to two different events). 
One then constructs the ratio
\begin{equation}
C(\Delta\phi)\equiv\frac{P_{\rm corr}(\Delta\phi)}{P_{\rm uncorr}(\Delta\phi)},
\end{equation}
This so-called ``mixed event'' technique enables the extraction of the 
physical correlations between the particles, eliminating the effects of 
an uneven detector acceptance. 
Neglecting nonflow correlations, one has in general 
\begin{equation}
\label{cdeltaphi}
C(\Delta\phi)=
\sum_{n=-\infty}^{+\infty} v_n^2 e^{in\Delta\phi},
\end{equation}
so that the Fourier expansion of the measured correlation function 
$C(\Delta\phi)$ simply yields the integrated flow $v_n$. 
Similarly, if one replaces $\phi_1$ with the azimuthal angle $\psi$ 
of a particle in a narrow phase space window, $v_n^2$ in 
Eq.~(\ref{cdeltaphi}) is replaced with $v'_nv_n$, where $v'_n$ 
is the differential flow of the particle of interest.

The values of $v_n$ and $v'_n$ obtained with this method coincide 
with the values we obtain from the cumulant of order 2, which are 
denoted by $v_n\{2\}$ and $v'_n\{2\}$ in this paper; 
while we do not need mixed events, we must apply correction factors
to our reconstructed $v_n\{2\}$ and $v'_n\{2\}$ if the detector 
does not have full azimuthal coverage, as explained at the end of 
Secs.\ \ref{s:int_flow} and \ref{s:diff_flow}. 
With the mixed-event technique, this correction is not 
required \cite{Lacey:2001}. 

The essential limitation of the two-particle correlation technique 
is that it is not possible to separate flow and nonflow correlations: 
therefore the results may be strongly biased by nonflow correlations,
which are eliminated in our method by the construction of higher order 
cumulants. 

The same limitation applies to the much more widely used subevent 
method \cite{Danielewicz:1985hn}. 
This methods involves two steps:
in each event, one constructs the flow vector (\ref{flowvector}) to 
estimate the orientation of the reaction plane $\Phi_R$, and a study 
of the azimuthal correlation between the flow vectors of two randomly 
chosen ``subevents'' yields the accuracy of this estimate (the so-called 
``reaction plane resolution''). This first step amounts to measuring 
the integrated flow. 
Then, in order to measure differential flow, one studies the azimuthal 
correlation between a single particle and the flow vector. 
Since the latter involves a summation over many particles, the 
correlation between a single particle and the flow vector is 
much stronger than the correlation between two single particles, 
which is probably the reason why this method is used so often. 
However, the relative weights of nonflow and flow correlations 
is the same \cite{Borghini:2001sa} as in the much simpler 
two-particle correlation technique discussed above, so that both 
suffer from the same limitations. 

Flow and nonflow correlations could in fact be distinguished in the 
subevent method, through a more detailed study of the azimuthal 
correlation between subevents. Indeed, flow and nonflow correlations 
yield a different shape of the distribution of the relative azimuthal 
angle $\Delta\Phi$ between subevents, as shown in 
\cite{Ollitrault:1995dy}. If correlations are only due to flow, 
then the distribution of $\Delta\Phi$ is universal, and 
depends only on the resolution $\chi\equiv v_n\sqrt{M}$ which 
characterizes the reaction plane resolution \cite{Ollitrault:1993ba}. 
A comparison of the calculated distribution with experimental 
data was recently done at energies of 250~MeV per nucleon 
\cite{Andronic:2001cx}. The agreement is perfect, which shows that 
the observed correlations are dominated by flow at these energies. 
To our knowledge, no such comparison has been carried out so far at 
ultrarelativistic energies.

\section{Results of Monte-Carlo simulations}
\label{s:mc}

We have performed various Monte-Carlo simulations to check the validity 
of the procedures explained in this paper.
In each simulation, $\N$ events are simulated; in a given 
event, the reaction plane $\Phi_R$ is chosen randomly, 
then ``pions'' and ``protons'' are generated 
according to the respective distributions
\begin{equation}
{dN\over d\phi}\propto 1+2v_1\cos(\phi-\Phi_R)+
2v_2\cos\left(2(\phi-\Phi_R)\right)
\end{equation}
and 
\begin{equation}
{dN'\over d\psi}\propto 1+2v'_1\cos(\psi-\Phi_R)+
2v'_2\cos\left(2(\psi-\Phi_R)\right)
\end{equation}
respectively.
We then reconstruct the integrated (Sec. \ref{s:mc_int}) and differential 
(Sec.\ \ref{s:mc_diff}) flows following the procedures exposed in 
Secs.\ \ref{s:integrated} and \ref{s:differential}. 

\subsection{Integrated flow}
\label{s:mc_int}

In the first set of  simulations, we generated $\N=10^5$ events with 
$M=200$ pions emitted with an integrated elliptic flow $v_2=6\%$, 
which we then tried to reconstruct. 
No integrated directed flow was simulated. 
We first assumed a perfect detector.  
The optimal cumulant order, 
defined by Eq.~(\ref{err-stat}), is $2k_{\rm opt}=4.2$. The 
estimate $v_2\{4\}$ derived from the fourth-order cumulant 
is thus likely to  give the best compromise between 
systematic and statistical errors, but we also calculated the 
estimates $v_2\{2\}$ and $v_2\{6\}$. 
\begin{table}[htbp]
\begin{center}
\begin{tabular}{|c|c|c|c|}
\hline
\cline{2-4}
\multicolumn{1}{l|}{}
& $v_2\{2\}$  & $v_2\{4\}$  & $v_2\{6\}$\\
\hline
full acceptance, $M=200$ 
&$6.01\pm 0.02$
&$6.03\pm 0.04$
&$6.08\pm 0.05$\\
\hline
full acceptance, $150<M<250$ 
&$6.01\pm 0.02$
&$6.11\pm 0.04$
&$6.13\pm 0.05$\\
\hline
nonflow correlations, $M=200$ 
&$9.21\pm 0.02$
&$6.16\pm 0.04$
&$6.19\pm 0.05$\\
\hline
``bow tie'' acceptance, $M=100$  
&$5.99\pm 0.04$
&$6.09\pm 0.12$
&$6.34{\scriptsize\matrix{+0.19\cr -0.22}}$\\
\hline
``bow tie'' acceptance, $75<M<125$  
&$5.85\pm 0.04$
&$5.40\pm 0.12$
&$5.74{\scriptsize\matrix{+0.19\cr -0.22}}$\\
\hline
\hline
\end{tabular}
\end{center}
\caption{Reconstruction of integrated elliptic flow. 
$\N=10^5$ events with $v_1=0$ and $v_2=6\%$ were generated in each simulation. 
The three columns give the values of $v_2$, in $\%$, 
reconstructed using cumulants to order $2$, 
$4$, and $6$, respectively. }
\label{MC1}
\end{table}

The results are presented in Table \ref{MC1}.
The error bars are statistical only. They are asymmetric for 
$v_2\{6\}$: this reflects the fact 
that the statistical fluctuations of $v_n\{2k\}$ are not Gaussian 
when the error is large, as explained in Appendix\ \ref{s:stat_int}. 
When working with a fixed multiplicity $M=200$, the  reconstructed 
values coincide with the theoretical value within expected statistical
errors. 
Note that the statistical error on $v_2\{6\}$ 
is only slightly larger than that on $v_2\{4\}$ in this case. 
When the multiplicity $M$ is randomly chosen between 150 and 250,
the reconstructed flow deviates from 
the theoretical value by more than two standard deviations, but the 
accuracy is still good, and moreover the statistical errors were 
calculated assuming a fixed detected multiplicity, while the real 
errors are probably larger.

In order to check the ability of our method to eliminate two-particle 
nonflow correlations, the latter were simulated by emitting particles 
in pairs, where both particles in a pair have the same azimuthal angle. 
$100$ pairs were emitted in each event, resulting in a multiplicity 
$M=200$. The reconstructed value with the cumulant of order 2 , i.e., 
without removing two-particle nonflow correlations, is 50\% too large; 
the standard methods would give similar results. 
On the other hand, the values obtained using higher order cumulants are 
much closer to the theoretical value; they are beyond statistical error 
bars, but this is not surprising since error bars were calculated assuming 
$M=200$ independent particles, while the effective multiplicity 
here is rather 100, which results in larger fluctuations. 

In summary, the results obtained so far show that $v_2\{4\}$ is 
to be preferred here: the statistical error on $v_2\{6\}$ is 
(slightly) larger, while nonflow correlations may give a large, 
uncontrolled, contribution to $v_2\{2\}$. 
\begin{center}
\begin{figure}[ht!]
\centerline{\includegraphics*[width=0.3\linewidth]{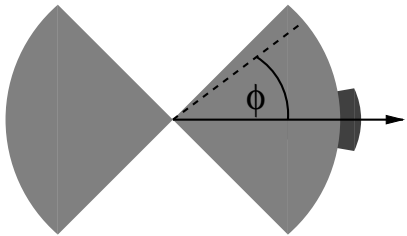}}
\medskip
\caption{Schematic picture of the PHENIX detector at RHIC. 
The shaded area indicates the azimuthal coverage of the detector, 
while the darker area at the right of the figure corresponds to a 
small time-of-flight detector with an extension of $45^\circ$, 
which can be used to measure the differential flow of identified 
particles.}
\label{fig:phenix}
\end{figure}
\end{center}
To test the validity of our acceptance corrections, we then did simulations 
with the ``bow-tie'' detector schematically represented in 
Fig.~\ref{fig:phenix}, which mimics the azimuthal acceptance of the 
PHENIX detector \cite{Lacey:2001}: particles are detected only within 
two quadrants of $90$ degrees each, with $100\%$ efficiency.
Since only half the particles are detected with this detector, the 
value of $M$ chosen in this simulation was half the value chosen above 
for a perfect detector. 
The optimal cumulant order is now $2k_{\rm opt}=4.5$, so that the fourth-order 
cumulant is still to be preferred. 
Since the azimuthal coverage is only partial, Eqs.~(\ref{c246})
relating the cumulant to the flow are no longer valid. 
In the case of elliptic flow, they are replaced by Eqs.~(\ref{c24accv2}). 
These formulas involve the Fourier coefficients $a_p$ of the acceptance 
function, defined by Eq.~(\ref{defap}). 
With the acceptance depicted in Fig.\ \ref{fig:phenix}, a simple 
calculation gives $a_2=2/\pi$ and $a_1=a_3=a_4=0$. 
Inserting these values into Eqs.~(\ref{c24accv2}), one finds
\begin{eqnarray}
\label{c24acc}
c_2\{2\} & \simeq & 0.518\, v_2\{2\}^2,\cr
c_2\{4\} & \simeq & -0.384\, v_2\{4\}^4.
\end{eqnarray}
An interesting feature of this bow-tie acceptance is that there is 
no interference between $v_1$ and $v_2$, because all $a_p$ with odd $p$
vanish. However, this is not the case in general (see 
Appendix \ref{s:acceptance}).
The results are given in Table \ref{MC1}.
Note that the estimate of statistical errors was 
done assuming a full acceptance, so they may 
be underestimated here. 
When the multiplicity $M$ is fixed, the reconstructed values agree 
with the theoretical values within eror bars. 
If the acceptance correction had not been applied, the reconstructed 
values would be below $5\%$. 

If $M$ is allowed to fluctuate, the discrepancy between reconstructed
and theoretical values is much larger than statistical errors, 
and also much larger than in the case of a full acceptance. 
As explained in Appendix\ \ref{s:cumulants}, 
a fluctuation $\delta M$ in multiplicity induces some errors, which 
scale like $\delta M^2$ and are comparatively larger when the acceptance 
is not good. 
In order to avoid this effect, the analysis can be done by selecting 
randomly a subset of the detected particles, with fixed multiplicity, 
as explained at the end of Sec.\ \ref{s:int_gene}. 
\begin{table}[htbp]
\begin{center}
\begin{tabular}{|c|c|c|}
\hline
\cline{2-3}
\multicolumn{1}{l|}{}
& $v_2=0$  & $v_2=6\,\%$ \\
\hline
full acceptance, $M=200$ 
&$2.01{\scriptsize\matrix{+0.21\cr -0.32}}$
&$2.19{\scriptsize\matrix{+0.21\cr -0.32}}$\\
\hline
``bow-tie'' acceptance, $M=200$  
&$1.87{\scriptsize\matrix{+0.21\cr -0.32}}$
&$1.88{\scriptsize\matrix{+0.21\cr -0.32}}$\\
\hline
\end{tabular}
\end{center}
\caption{Reconstruction of a theoretical directed flow $v_1=2\,\%$. 
$\N=5\times 10^5$ events were 
generated in each simulation, with $M=200$ detected particles 
in each event. 
The table gives the values of $v_1\{4\}$ in $\%$. }
\label{MC2}
\end{table}

We then performed a second set of simulations, with both 
directed and elliptic flow, in order to test possible interferences 
between the two. Such uncontrolled interferences were the main limitation of 
our previous cumulant method \cite{Borghini:2001sa}, but are avoided
here, as explained in Sec.\ \ref{s:ourmethod}. 
Inteference of a different kind may in fact still occur when the 
detector has partial azimuthal coverage, as explained in 
Appendix \ref{s:accint}, but they are under control. 
Events were generated with a small directed flow $v_1=2\%$, 
and with an elliptic flow $v_2=0$ or $v_2=6\%$. The goal 
was to reconstruct $v_1$. 
Since $2\%$ is a very small value, a larger number of events were 
generated than in the previous set of simulations. 
With the values of $\N$ and $M$  given in Table \ref{MC2},
the optimal cumulant order is still 4: all the results presented 
in Table \ref{MC2} are reconstructed from the fourth-order cumulant.
In the case of the bow-tie acceptance, 
the relation between $v_1\{4\}$ and the cumulant, Eq.~(\ref{c4}), 
must be replaced with (\ref{c4accv1}), which gives 
\begin{equation}
\label{c14acc}
c_1\{4\}\simeq -2.785\, v_1\{4\}^4. 
\end{equation}
The reconstructed value of $v_1$ is always compatible with the theoretical 
value $v_1=2\%$ within the statistical error, even with the bow-tie 
acceptance. 
This is a major improvement on our previous method 
\cite{Borghini:2001sa}, where the reconstruction of $v_1$ would have 
been impossible if  $v_2=6\%$, 
even in the case of a perfect detector. 

\subsection{Differential flow}
\label{s:mc_diff}

We then performed various simulations to test our reconstruction 
of the differential flow. 
In a first set of simulations (see Table\ \ref{MC3}), only elliptic 
flow was generated.
Events were generated with $v'_2=v_2=6\%$, and we tried to reconstruct 
$v'_2$. 
As explained in Sec.\ \ref{s:differential}, the first step 
in the analysis is the reconstruction of $v_2$. 
Since no integrated directed flow $v_1$ is present, $v'_2$ 
can then be reconstructed only with respect to $v_2$: 
this is denoted by $v'_{2/2}$ in Sec.\ \ref{s:differential}. 
One can reconstruct it from the lowest order cumulant. 
This yields the estimate $v'_{2/2}\{2\}$, equivalent 
to the standard flow analysis. One can also reconstruct 
the higher order cumulant $v'_{2/2}\{4\}$, which eliminates
 nonflow correlations between the proton and  the pions. 

\begin{table}[htbp]
\begin{center}
\begin{tabular}{|c|c|c|c|}
\hline
pions & protons & $v'_{2/2}\{2\}$  & $v'_{2/2}\{4\}$  \\
\hline
full acceptance, $M=200$ & full acceptance
&$5.96\pm 0.15$
&$6.10\pm 0.44$\\
\hline
``bow-tie'' acceptance, $M=100$  & ``bow-tie'' acceptance
&$6.07\pm 0.20$
&$5.83\pm 0.99$\\
\hline
``bow tie'' acceptance, $M=100$ 
& 45$^\circ$ acceptance
&$6.02\pm 0.20$
&$5.71\pm 0.99$\\
\hline
\end{tabular}
\end{center}
\caption{Reconstruction of differential 
elliptic flow with respect to integrated elliptic flow, i.e., 
$v'_{2/2}$. 
In each simulation, $\Np=5\times 10^5$ protons were generated with 
$v'_2=6\%$ and $v'_1=0$.
The integrated flow was also $v_2=6\%$ and $v_1=0$. 
The last two columns give the values of $v'_2$  (in $\%$) reconstructed 
from the cumulants to 
order $2$ and  $4$, respectively. }
\label{MC3}
\end{table}

In the first simulation, we assumed that the detector 
was perfect, both for integrated and for differential flow. 
With the values of the parameters given in Table\ \ref{MC3}, 
the optimal cumulant order, defined by Eq.~(\ref{err-statd}), 
is $(2k+m+1)_{\rm opt}=3.5$, i.e., $v_{2/2}\{4\}$ is to be preferred. 
Nevertheless, we also include the lowest order estimates
$v_{2/2}\{2\}$. 
Since no nonflow correlations are generated between the 
proton and the pions, this estimate should be good. 
Indeed, both estimates are found to be in very good agreement 
with the theoretical value.

We then turned to the case when the detector 
no longer has a perfect acceptance. 
In this case, corrections must be applied, as explained in 
Appendix \ref{s:accdiff}. 
For the integrated flow, we used the ``bow-tie'' detector 
of Fig. \ref{fig:phenix}. 
As in Sec.\ \ref{s:mc_int}, the multiplicity of each event with this 
detector is $M=100$, instead of $M=200$ with the full acceptance. 
This modification does not affect the optimal cumulant order, which 
is still 4.
For differential flow, two different detectors were considered. 
First, the same bow-tie acceptance as for pions, i.e., the relevant 
Fourier coefficients are $a'_2=2/\pi$ and $a'_1=a'_3=a'_4=0$.
In this case, the relations between the cumulants and flow, 
Eqs.~(\ref{d24accv'2}), are
\begin{eqnarray}
d_{2/2}\{2\} & \simeq & 0.595\, v_{2}\, v'_{2/2}\{2\}, \cr
d_{2/2}\{4\} & \simeq & -0.406\, v_2^3\, v'_{2/2}\{4\}.
\end{eqnarray}
Secondly, we assumed the protons were detected with a small $45^{\rm o}$ 
detector (see Fig. \ref{fig:phenix}). 
The Fourier coefficients of its acceptance function $A'(\psi)$ are 
$a'_0=1$, $a'_{p \geq 1}=8\sin(p\pi/8) / (p\pi)$, which yields the 
relations
\begin{eqnarray}
d_{2/2}\{2\} & \simeq & 0.337\, v_2\, v'_{2/2}\{2\}, \cr
d_{2/2}\{4\} & \simeq & -0.181\, v_2^3\, v'_{2/2}\{4\}.
\end{eqnarray}
Note that the last correction is more than a factor of 5, 
compared to a full acceptance. 
In all cases, the results (Table \ref{MC3}) are in 
excellent agreement with the theoretical value. This shows that with 
our method, it is possible to measure differential flow with any detector 
provided the relevant acceptance corrections are performed. 

\begin{table}[htbp]
\begin{center}
\begin{tabular}{|c|c|c|c|c}
\hline
pions & protons & $v'_{2/1}\{3\}$ &
 $v'_{2/2}\{2\}$  & $v'_{2/2}\{4\}$  \\
\hline
full acceptance, $M=200$ & full acceptance
&$6.47\pm 0.32$
&$5.81\pm 0.15$
&$6.03\pm 0.44$
\\
\hline
``bow-tie'' acceptance, $M=100$  & ``bow-tie'' acceptance
&$5.08\pm 0.54$
&$5.80\pm 0.20$
&$5.51\pm 0.99$\\
\hline
``bow tie'' acceptance, $M=100$ 
& 45$^\circ$ acceptance
&$5.20\pm 0.54$
&$6.81\pm 0.20$
&$7.95\pm 0.99$\\
\hline
\end{tabular}
\end{center}
\caption{Same as Table \ref{MC3}, except that a directed flow 
$v_1=v'_1=6\%$ was generated in addition to the elliptic flow. 
The third column gives the differential elliptic flow recontructed 
with respect to integrated directed flow, to lowest order.}
\label{MC4}
\end{table}

We then did a second set of simulations. 
The only difference with the first set is that 
directed flow is also generated, with the same magnitude 
as elliptic flow, i.e., $v'_1=v_1=6\%$. 
In this case, it is also possible to measure differential elliptic 
flow with respect to integrated directed flow, $v'_{2/1}$. 
We recontructed only the lowest order estimate $v'_{2/1}\{3\}$, 
which is also given by the standard flow analysis. Note that 
it is already insensitive to two-particle nonflow correlations. 
In the case when the bow-tie acceptance applies to both integrated 
and differential flow, Eq.~(\ref{d3accv'1}) does not give any 
correction, so that Eq.\ (\ref{d_2n/n2}) still holds. 
When protons are measured with a small $45^\circ$ detector, on the 
other hand, Eq.~(\ref{d3accv'1}) gives 
\begin{equation}
d_{2/1}\{3\} \simeq 1.258\, v_1^2 v'_{2/1}\{3\}.
\end{equation}
Results are given in Table\ \ref{MC4}. 
With a full acceptance, the reconstruction of both $v'_{2/1}$ and 
$v'_{2/2}$ is good, within the expected error bars. 
With a partial acceptance, however, discrepancies are much larger.
In particular, the value of $v'_{2/2}$ with a $45^\circ$ acceptance 
differs by four standard deviations from the theoretical value. 

\section{Conclusions}
\label{s:conclusions}

We have presented a new, general method for analyzing the flow. 
This method can be used by all heavy-ion experiments which implement 
the standard analysis of Danielewicz and 
Odyniec~\cite{Danielewicz:1985hn}, since it uses the same input,
namely the azimuthal angles of outgoing particles. 
As in the standard analysis, one proceeds in two steps. 
One first reconstructs the average value of the flow over phase 
space, with an appropriate weight: 
this ``integrated flow'' $v_n$ is related to the reaction plane resolution 
in the standard analysis. 
One may then perform detailed analyses in narrower phase-space
windows, i.e., measure ``differential flow'' $v'_n$.
The strong points of our method are the following:
\begin{itemize}
\item{It  systematically eliminates azimuthal correlations which are 
not due to flow.}
\item{One obtains several different estimates of $v_n$ and $v'_n$; 
comparison between these estimates provides a useful consistency check.}
\item{The detectors need not have full azimuthal coverage.}
\end{itemize}
Let us comment on these three points in more detail. 

Nonflow correlations are eliminated by means of a cumulant 
expansion of multiparticle azimuthal correlations. 
This cumulant expansion applies to both integrated and differential
flow. In the case of integrated flow, one constructs a series of
cumulants $c_n\{2k\}$, where $k$ is an arbitrary positive integer. 
$c_n\{2k\}$ involves correlations between $2k$ particles, but is 
insensitive to correlations involving less than $2k$ particles. 
Now, flow is by definition a collective phenomenon, which induces 
correlations between all the produced particles: as $k$ increases, 
its relative contribution to $c_n\{2k\}$ also increases. 

Using the cumulant $c_n\{ 2k\}$, one obtains an estimate 
of $v_n$, denoted by $v_n\{2k\}$ in this paper. The lowest 
order $v_n\{2\}$ corresponds to the standard flow analysis:
it coincides with $v_n$ if all correlations are due to flow. 
The nice feature of our method is that higher order cumulants 
provide us with other, independent estimates of $v_n$, which can be used 
as a consistency check. 
The reader might believe that constructing cumulants of six-particle 
correlations requires a huge statistics, and is practically
impossible. 
This is not true. Statistical errors on higher order 
estimates $v_n\{2k\}$ are discussed thoroughly in 
Appendix \ref{s:statistical}. They strongly depend on the 
strength of the flow itself, and more precisely on the parameter 
$\chi\equiv v_n\sqrt{M}$, which characterizes the resolution 
of the reaction plane reconstruction, $\Delta\Phi_R$, in the
standard analysis. If $\chi=1$, corresponding to
a resolution $\Delta\Phi_R=45^\circ$, which is achieved in many 
experiments, the statistical error on both $v_n\{4\}$ and $v_n\{6\}$ is 
only 60\% higher than the error on $v_n\{2\}$, as can be seen 
in Fig.\ \ref{fig:stat}. 

The same discussion can be repeated for differential flow. 
In the case of differential elliptic flow $v'_2$, however, 
the standard analysis also provides us with two different 
estimates of this quantity, depending on whether the reaction 
plane is measured using directed ($v'_{2/1}$) or elliptic ($v'_{2/2}$)
flow. The first of these two estimates, $v'_{2/1}$, which involves 
correlations between three particles, is insensitive to two-particle
nonflow correlations, even in the standard analysis. 
In this paper, we obtain several estimates for both $v'_{2/1}$ and 
$v'_{2/2}$, which can be used as further consistency checks. 

Finally, we have also included a detailed discussion of acceptance 
corrections, which allow us to work even if the detector has 
only partial azimuthal coverage. 
One may argue that the corresponding formulas, given in 
Appendix\ \ref{s:acceptance}, are very heavy. 
However, these formulas are required only if the acceptance is far 
from isotropic: corrections corresponding to weak inhomogeneities of 
the acceptance are automatically taken care of by the cumulant 
expansion itself. 
Furthermore, the correction factors need only be calculated once 
for a given detector, and it is a simple calculation. 
When they are applied, accurate results can be obtained even 
with a very poor acceptance, as illustrated by the simulations 
performed in Sec.\ \ref{s:mc}. 
Although several methods have been proposed to correct for detector 
inefficiencies in the standard analysis \cite{Poskanzer:1998yz}, we 
do not know of any systematic study of their limits.

\section*{Acknowledgements}

N.~B.\ thanks the Nuclear Chemistry group at Stony Brook for their 
hospitality, and in particular R.~Lacey and N.~N.~Ajitanand for 
challenging suggestions.
We thank E.~Guitter, A.~Poskanzer, R.~Snellings, Aihong Tang, 
S.~Voloshin, and A.~Wetzler for discussions. 

N.\ B.\ acknowledges support from the ``Actions de Recherche Concert\'ees'' 
of ``Communaut\'e Fran\c{c}aise de Belgique'' and IISN-Belgium.

\appendix
\section{Cumulants of multiparticle azimuthal correlations}
\label{s:cumulants}

In this Appendix, we justify the definition of the cumulants 
from the generating function (\ref{defc0}), 
by discussing  a few examples. 

We begin with the second order cumulant 
$c_n\{2\}=\cumul{e^{in(\phi_1-\phi_2)}}$. 
Inserting Eq.~(\ref{expansionnewG0}) in expression (\ref{defc0}) and 
expanding to order $zz^*$, we obtain from Eq.~(\ref{defc1}) 
\begin{equation}
\label{appa1}
\cumul{e^{in(\phi_1-\phi_2)}}= 
\frac{M-1}{M} \left(\mean{e^{in(\phi_1-\phi_2)}} -
  \mean{e^{in\phi_1}}\mean{e^{-in\phi_2}}\right). 
\end{equation}
If particles are uncorrelated, i.e. 
$\mean{e^{in(\phi_1-\phi_2)}}=\mean{e^{in\phi_1}}\mean{e^{-in\phi_2}}$.
the cumulant vanishes, as it should.  
In the limit when $M\gg 1$, one recovers definition (\ref{defc2}). 

The standard definition of the cumulants is 
${\cal C}_n(z)=\ln\mean{G_n(z)}$, rather than  
Eq.~(\ref{defc0}) \cite{vanKampen}. 
However, with this definition we would obtain, instead of 
Eq.~(\ref{appa1}),
\begin{eqnarray}
\label{appa2}
\cumul{e^{in(\phi_1-\phi_2)}}
&=& \frac{M-1}{M} \mean{e^{in(\phi_1-\phi_2)}} -
  \mean{e^{in\phi_1}}\mean{e^{-in\phi_2}}\cr
&=& \frac{M-1}{M} \left(\mean{e^{in(\phi_1-\phi_2)}} -
  \mean{e^{in\phi_1}}\mean{e^{-in\phi_2}}\right) 
+\frac{1}{M}  \mean{e^{in\phi_1}}\mean{e^{-in\phi_2}}. 
\end{eqnarray}
If particles are uncorrelated, the last term in the 
equation remains if the acceptance is not perfect. 
This may result in large errors when the flow is small. 

In Sec.\ \ref{s:int_gene}, we also considered the possibility 
that the multiplicity $M$ be not strictly the same for 
all events. If it fluctuates around an average value $\mean{M}$, 
then the generating function (\ref{expansionnewG0}) must also
be averaged over $M$, and 
$M$ must be replaced by $\mean{M}$ in the definition 
of the cumulants,  Eq.~(\ref{defc0}). 
Equation (\ref{appa1}) is then replaced by 
\begin{equation}
\label{appa3}
\cumul{e^{in(\phi_1-\phi_2)}}= 
\mean{\frac{M-1}{M}} 
\mean{e^{in(\phi_1-\phi_2)}} -
\left(1-{1\over\mean{M}}\right)\mean{e^{in\phi_1}}\mean{e^{-in\phi_2}}. 
\end{equation}
The magnitude of the fluctuations of $M$ can be characterized by 
their standard deviation $\delta M^2\equiv \mean{M^2}-\mean{M}^2$. 
If $\delta M\ll \mean{M}$, then 
$\mean{1/M}\simeq 1/\mean{M}+\delta M^2/\mean{M}^3$, and 
Eq.~(\ref{appa3}) becomes  
\begin{equation}
\label{appa4}
\cumul{e^{in(\phi_1-\phi_2)}}= 
\mean{\frac{M-1}{M}} 
\left(\mean{e^{in(\phi_1-\phi_2)}} -
  \mean{e^{in\phi_1}}\mean{e^{-in\phi_2}}\right)-
{\delta M^2\over\mean{M}^3}\mean{e^{in\phi_1}}\mean{e^{-in\phi_2}},
\end{equation}
to be compared with Eq.~(\ref{appa1}). 
If $\delta M\ll M$ and $M\gg 1$, the correction should be negligible. 

This is why the factor $1/M$ associated with each particle in the generating 
function (\ref{newG0}) is important. 
If this factor had not been included, the coefficient in front of $zz^*$ 
in ${\cal C}_n(z)$ would be
\begin{eqnarray}
\label{appa5}
\mean{M(M-1)} \mean{e^{in(\phi_1-\phi_2)}}  &-&
\mean{M}(\mean{M}-1)
\mean{e^{in\phi_1}}\mean{e^{-in\phi_2}} \cr
&=& 
\mean{M(M-1)} \left(\mean{e^{in(\phi_1-\phi_2)}} -
\mean{e^{in\phi_1}}\mean{e^{-in\phi_2}}\right)
+\delta M^2\mean{e^{in\phi_1}}\mean{e^{-in\phi_2}}.
\end{eqnarray}
In this equation, the second term in the r.-h.~s. appears with a 
coefficient of order $\delta M^2/\mean{M}^2$ with respect to the 
coefficient of the first term. 
This should be compared with Eq.~(\ref{appa4}), where 
the coefficient of the second term is of magnitude 
$\delta M^2/\mean{M}^3$, i.e., much smaller. 
Thus the weight $1/M$ in Eq.~(\ref{newG0}) minimizes the 
effects of a fluctuating multiplicity. 

We now turn to the expression of the fourth-order cumulant. 
For simplicity, we restrict the discussion to a perfect 
detector, and to a fixed multiplicity $M$. 
Then, the generating function $\mean{G_n(z)}$ is 
\begin{equation}
\mean{G_n(z)} = 1 + \frac{M-1}{M} |z|^2\mean{e^{in(\phi_1-\phi_2)}} +
\frac{(M-1)(M-2)(M-3)}{4M^3} |z|^4
\mean{e^{in(\phi_1+\phi_2-\phi_3-\phi_4)}} + \ldots 
\end{equation}
Inserting this expression in Eq.~(\ref{defc0}) and expanding 
to order $|z|^4$, one obtains 
\begin{equation}
\frac{1}{4} \cumul{e^{in(\phi_1+\phi_2-\phi_3-\phi_4)}}  
=\frac{M-1}{4\,M}\left[\frac{(M-2)(M-3)}{M^2} 
  \mean{e^{in(\phi_1+\phi_2-\phi_3-\phi_4)}} -
  2\frac{(M-1)^2}{M^2} \mean{e^{in(\phi_1-\phi_2)}}^2 \right] 
\end{equation}
which gives Eq.~(\ref{defc4}) in the limit of large $M$. 

More generally, the cumulants $\langle\langle
\exp[in(\phi_1+\cdots+\phi_k-\phi_{k+1}-\cdots-\phi_{k+l})]\rangle\rangle$
derived from the generating equation (\ref{defc0}) coincide with the 
quantities of physical interest, that is, only the ($k+l$)-particle direct 
correlations when there is no flow, up to subleading terms of relative
magnitude ${\cal O}(1/M)$.

\section{Interpolation formulas}
\label{s:interpolation}

In this Appendix, we give interpolation methods to compute numerically 
the cumulants from their generating functions. 
Generally, one wishes to reconstruct $k_i$ cumulants for integrated 
flow, i.e., $c_n\{2k\}$ for $k=1,\cdots,k_i$,  and $k_d$ cumulants 
for differential flow, i.e., $d_{mn/n}\{2k+m+1\}$ for 
$k=0,\cdots,k_d-1$. 
Typical values are $k_i=3$, $m=1,2$ and $k_d=2$. 
One thus obtains three independent estimates of integrated flow $v_n$, 
and two estimates of each $v'_{mn/n}$. 

The cumulants are defined as coefficients in the power-series 
expansions of the generating functions ${\cal C}_n(z)$ 
[Eq.~(\ref{defc1})] and ${\cal D}_{mn/n}(z)$ [Eq.~(\ref{defcm1})]. 
To extract the cumulants numerically, one first computes the 
generating functions at the points $z_{p,q}=x_{p,q}+iy_{p,q}$ with 
\begin{eqnarray}
\label{defxy}
x_{p,q}&\equiv& r_0\sqrt{p}\,\cos\left({2\,q\,\pi\over q_{\rm max}}\right)\cr
y_{p,q}&\equiv& r_0\sqrt{p}\,\sin\left({2\,q\,\pi\over q_{\rm max}}\right)
\end{eqnarray}
for $p=1,\cdots,k_{\rm max}$ and $q=0,\cdots,q_{\rm max}-1$. 
These equations define a set of $k_{\rm max}\times q_{\rm max}$ 
points in the complex plane, where $k_{\rm max}$ and $q_{\rm max}$ 
must satisfy $k_{\rm max}\ge k_i, k_d$ and 
$q_{\rm max}>2k_i,2(k_d+m-1)$. 
With the values of $k_i$, $m$, and $k_d$ given above, one can choose 
for instance $k_{\rm max}=3$ and $q_{\rm max}=7$ or 8. 

The number $r_0$ in Eqs.~(\ref{defxy}) is in principle a small number, 
since we are interested in the behaviour of the generating functions 
near the origin. 
If $r_0$ is too small, however, large numerical errors occur. 
In practice, the numerical simulations presented in Sec.\ \ref{s:mc} were 
done with double precision numbers (16 digits), and with the 
value $r_0=1.5$. 
A much smaller value was given in \cite{Borghini:2001sa}. 
This is due to a rescaling of the variable $z$ by a factor of 
$\sqrt{M}$, as discussed in Sec.~\ref{s:ourmethod} 
[see Eq.~(\ref{newweight})]. 
In any case, the interpolation should be done with two different 
values of $r_0$ in order to check the stability of the results.

\subsection{Integrated flow}
\label{s:interpoli}

We denote by $C_{p,q}$ the values of the generating function 
${\cal C}(z)$ evaluated at the points (\ref{defxy}):
\begin{equation}
C_{p,q}  \equiv  {\cal C}_n( z_{p,q}).  
\end{equation}
The cumulants $c_n\{ 2k\}$ correspond to the terms with $k=l$ 
in the power-series expansion (\ref{defc1}). In order to eliminate 
terms with $k\not=l$, one averages over the phase of z:
\begin{equation}
C_{p}\equiv {1\over q_{\rm max}}\sum_{q=0}^{q_{\rm max}-1}C_{p,q}.
\end{equation}
Then, the $C_p$, with $p=1,\dots,k_i$, 
are related to the cumulants 
$c_n\{2k\}$ with $k=1,\dots,k_i$ 
by the following linear system of $k_i$ equations:
\begin{equation}
\label{syst-Cp}
C_p=\sum_{k=1}^{k_i} 
{(r_0 \sqrt{p})^{2k}\over (k!)^2}\,c_n\{2 k\}, \qquad 1 \leq p \leq k_i.
\end{equation}
The solution of this system for $k_i=3$ reads 
\begin{eqnarray}
\label{into6}
c_n\{2\}&=&{1\over r_0^2}\,
\left(3\, C_1-{3\over 2}C_2+{1\over 3}C_3\right)
,\cr
c_n\{4\}&=&{2\over r_0^4}\,
\left(-5\, C_1+4\,C_2-\,C_3\right)
,\cr
c_n\{6\}&=&{6\over r_0^6}\,
\left(3\, C_1-3\,C_2+C_3\right).
\end{eqnarray}
Solving Eqs.~(\ref{syst-Cp}) with a larger value of $k_i$ provides 
more accurate values of the first three cumulants, as well as higher 
order cumulants $c_n\{2k\}$.

\subsection{Differential flow}
\label{s:interpold}

The generating function ${\cal D}_{mn/n}(z)$ is complex. 
From definition (\ref{defcm}), its real and imaginary parts at the 
points $z_{p,q}$ are 
\begin{eqnarray}
X_{p,q}&\equiv& {\Re}\left[{\cal D}_{mn/n}(z_{p,q})\right]=
\frac{\mean{\cos(mn\psi)\,G_n(z_{p,q})}}{\mean{G_n(z_{p,q})}}\cr
Y_{p,q}&\equiv& {\Im}\left[{\cal D}_{mn/n}(z_{p,q})\right]=
\frac{\mean{\sin(mn\psi)\,G_n(z_{p,q})}}{\mean{G_n(z_{p,q})}}.
\end{eqnarray}
The cumulants $d_{mn/n}\{2k+m+1\}$, defined in Eq.~(\ref{defdk}), 
are the real parts of the terms proportional to $z^{*k}z^{k+m}$ 
in the power-series expansion (\ref{defcm1}). 
In order to isolate these terms, one multiplies 
${\cal D}_{mn/n}(z)$ by $z^{*m}$, takes the real part,  
and averages over angles:
\begin{equation}
D_{p}\equiv {\left (r_0\sqrt{p}\right)^m 
\over q_{\rm max}}\sum_{q=0}^{q_{\rm max}-1}
\left[\cos\left(m\frac{2\,q\,\pi}{q_{\rm max}}\right) \, X_{p,q}+
\sin\left(m\frac{2\,q\,\pi}{q_{\rm max}}\right) \,Y_{p,q}\right]. 
\end{equation}
The values of $D_{p}$ for $p=1,\dots, k_d$ are related to the cumulants 
$d_{mn/n}\{2k+m+1\}$ with $k=0,\dots,k_d-1$ by the following linear 
system of $k_d$ equations:
\begin{equation}
D_{p}=\sum_{k=0}^{k_d}
{(r_0 \sqrt{p})^{2(k+m)}\over k!(k+m)!}\, d_{mn/n}\{2k+m+1\}, 
\quad 1 \leq p \leq k_d.
\end{equation}
For $k_d=2$ and $m=1$,  the solution of this system is 
\begin{eqnarray}
d_{n/n}\{2\}
&=&{1\over r_0^2}\left(2\, D_1-{1\over 2}\, D_2\right),\cr
d_{n/n}\{4\}
&=&{1\over r_0^4}\left( -2\, D_1+ D_2\right),
\end{eqnarray}
while for $k_d=2$ and $m=2$, 
\begin{eqnarray}
d_{2n/n}\{3\}
&=&{1\over r_0^4}\left(4\, D_1-{1\over 2}\, D_2\right),\cr
d_{2n/n}\{5\}
&=&{1\over r_0^6}\left( -6\,D_1+{3\over 2}\, D_2\right).
\end{eqnarray}

\section{Acceptance corrections}
\label{s:acceptance}

We derive here the relations between the cumulants and the flow 
when the detector has only partial azimuthal coverage. 
As explained in Sec.\ \ref{s:int_flow}, we assume that the class 
of events used in the flow analysis (usually corresponding to a given 
centrality cut) is selected by means of a detector which has, at 
least approximately, full azimuthal coverage, so that we may assume 
that the probability distribution of $\Phi_R$ is uniform. 

We wish to recall that the first step in our accounting for acceptance 
inhomogeneities is the choice of ``nonisotropic'' cumulants, as 
explained in Sec.\ \ref{s:int_gene}: while the simple cumulant 
deduced from Eq.\ (\ref{defc4bis}) is valid for a perfect detector, a 
more general definition is that deduced from Eqs.\ (\ref{defc0}) and 
(\ref{defc1}). 
For ``almost'' perfect detectors, this first step should be enough, 
and the following results over-sophisticated.

\subsection{Integrated flow}
\label{s:accint}

Let us describe the detector characteristics (acceptance and efficiency)
by a real-valued function $A(\phi)$, which is the probability that 
a particle emitted at angle $\phi$ be detected. 
We choose to normalize this function according to 
$\int_0^{2\pi}A(\phi)d\phi=2\pi$. 
The function $A(\phi)$  can be expanded into a Fourier series:
\begin{equation}\label{Aphi}
A(\phi) = \sum_{p=-\infty}^{+\infty} a_p\, e^{ip\phi},
\end{equation}
where the $a_p$ coefficients, Eq.~(\ref{defap}), satisfy $a_{-p}=a_p^*$, 
and $a_0=1$ due to the normalization choice. 
For a perfect detector, $a_{p \neq 0}=0$.

The distribution of outgoing particles in a given collision is 
\begin{equation}\label{Pphi}
P(\phi-\Phi_R) \propto \sum_{p=-\infty}^{+\infty} v_p e^{ip(\phi-\Phi_R)},
\end{equation}
where $\Phi_R$ is the reaction plane azimuth of the collision, $v_0=1$, 
and the $v_p=v_{-p}$ are real valued.

We now evaluate the generating function of azimuthal 
correlations $\mean{G_n(z)}$, with $G_n(z)$ defined by 
Eq.~(\ref{newG0}), and compute the cumulants 
as a function of the flow coefficients $v_p$ and 
the acceptance coefficients $a_p$.

For a given orientation of the reaction plane, the average value 
of $e^{-in\phi}$ for a particle seen in the detector is 
\begin{equation}\label{eq3}
\mean{e^{-in\phi}|\Phi_R} = 
\frac{\displaystyle\int_0^{2\pi}\! d\phi\, e^{-in\phi} A(\phi) P(\phi-\Phi_R)}{
\displaystyle\int_0^{2\pi}\! d\phi\, A(\phi) P(\phi-\Phi_R)} = 
\frac{\displaystyle \sum_{p=-\infty}^{+\infty} a_{p+n} v_p e^{ip\Phi_R}}{
\displaystyle \sum_{p=-\infty}^{+\infty} a_{p} v_p e^{ip\Phi_R}}, 
\end{equation}
where $n$ is the harmonic one wants to measure. The denominator 
is the probability that a pion be detected, which depends on 
$\Phi_R$ if there is flow. 

To obtain the cumulants to leading order in the flow coefficients $v_p$ 
with $p\not=0$, one can linearize this expression in $v_p$ with $p\not=0$:
\begin{equation}\label{eq4}
\mean{e^{-in\phi}|\Phi_R} = 
a_n+\sum_{p\not= 0} (a_{p+n}-a_n a_p) v_p e^{ip\Phi_R}.
\end{equation}
In this expression, the term proportional to $a_na_p$ comes from the 
denominator of Eq.~(\ref{eq3}). It reflects the $\Phi_R$-dependence 
of the probability that a pion be detected.

The average value (\ref{eq4}) can then be introduced in the generating 
function $\mean{G_n(z)}$. We neglect nonflow correlations for 
simplicity. Then, the angles of the particles are statistically independent 
and we obtain
\begin{eqnarray}
\label{genfunc}
\mean{G_n(z)|\Phi_R} &=& \left[ 1 + \frac{z}{M}
\left( a_n+\sum_{p\not= 0} (a_{p+n}-a_n a_p)\, v_p\, e^{ip\Phi_R}\right)
+ {\rm c.c.} \right]^M \cr
&\simeq & e^{z\,a_n+{\rm c.c.}}
\exp\left[z\,
\sum_{p\not= 0}(a_{p+n}-a_n a_p)\, v_p\, e^{ip\Phi_R}
+ {\rm c.c.} \right].
\end{eqnarray}
In the second identity, we have assumed that $M$ is large, so that 
$(1+x/M)^M\simeq \exp x$.
Equation (\ref{genfunc}) can be compared to Eq.~(\ref{meanG0}), 
to which it reduces when the acceptance is perfect. 
When it is not the case, the generating function depends in general on 
harmonics $v_p$ with $p\not= n$. 

The generating function must still be averaged over $\Phi_R$, as in 
Eq.~(\ref{avphiR}). 
The first term $\exp(z a_n+z^*a_n^*)$ is independent of $\Phi_R$ and 
factors out. 
This term does not contribute to the cumulants of order 2 and higher, 
since it gives a linear contribution to the generating function of the 
cumulants (\ref{defc0}), which is $\ln \mean{G_n(z)}$ in the large $M$ 
limit. 

In the general case, there is no simple analytic expression for the 
average over $\Phi_R$ of $\mean{G_n(z)}$. To obtain the cumulant at a 
given order, one must expands Eq.~(\ref{genfunc}) to the desired order, 
and then integrate over $\Phi_R$. 
Then Eq.~(\ref{defc1}) yields the cumulants.

Keeping only the first harmonics $v_1$ and $v_2$, which correspond to 
the terms $p=\pm 1,\pm 2$ in Eq.~(\ref{eq4}), one finally obtains for 
$n=1$ (i.e., for a measurement of directed flow) 
\begin{mathletters}
\label{c24accv1}
\begin{equation}
\label{c2accv1}
c_1\{2\}  =   \left[(1-|a_1|^2)^2 + |a_2-a_1^2|^2\right]v_1^2 
 + \left[|a_1-a_2a_1^*|^2 + |a_3-a_1a_2|^2\right] v_2^2 
\end{equation}
Both harmonics interfere, and one cannot measure $v_1$ and $v_2$ 
independently. 
Similarly, the cumulant to order 4 becomes 
\begin{eqnarray}
\label{c4accv1}
c_1\{4\} &=&  - \left[(1-|a_1|^2)^4 + 4\,(1-|a_1|^2)^2 |a_2-a_1^2|^2
+|a_2-a_1^2|^4\right]v_1^4 \cr
& & - \left[|a_1-a_2a_1^*|^4 + 4\,|a_1-a_2a_1^*|^2  |a_3-a_1a_2|^2 
+|a_3-a_1a_2|^4 
\right] v_2^4. 
\end{eqnarray}
\end{mathletters}
For $n=2$ (integrated elliptic flow), the first two cumulants are 
\begin{mathletters}
\label{c24accv2}
\begin{equation}
\label{c2accv2}
c_2\{2\}=   \left[|a_1-a_2a_1^*|^2 + |a_3-a_1a_2|^2\right]v_1^2 
 + \left[(1-|a_2|^2)^2 + |a_4-a_2^2|^2\right] v_2^2. 
\end{equation}
and 
\begin{eqnarray}
\label{c4accv2}
c_2\{4\}& = &  -\left[|a_1-a_2a_1^*|^4 + 4\,|a_1-a_2a_1^*|^2 
|a_3-a_1a_2|^2+|a_3-a_1a_2|^4 \right]v_1^4 \cr
& & -\left[ (1-|a_2|^2)^4 + 4\,(1-|a_2|^2)^2  
|a_4-a_2^2|^2+ |a_4-a_2^2|^4 \right] v_2^4. 
\end{eqnarray}
\end{mathletters}
If the acceptance is perfect, Eqs.~(\ref{c24accv1}) and (\ref{c24accv2}) 
reduce to Eqs.~(\ref{c2}) and (\ref{c4}), as they should. 
In the general case, 
Equations (\ref{c2accv1}) and (\ref{c2accv2}) represent a linear system 
of equations, which can easily be solved to give $v_1$ and $v_2$ (or more 
precisely, $v_1\{2\}$ and $v_2\{2\}$) as a function of the cumulants. 
Similarly, Eqs.~(\ref{c4accv1}) and (\ref{c4accv2}) can be solved 
to obtain $v_1\{4\}$ and $v_2\{4\}$ as functions of the cumulants. 

We do not include the expressions of the cumulants $c_1\{6\}$ and 
$c_2\{6\}$. In addition to the expected terms in $v_1^6$ and $v_2^6$, 
they involve a term proportional to $v_1^4v_2^2$. 

Note that in going from Eq.\ (\ref{eq3}) to Eq.\ (\ref{eq4}), 
we have kept only the leading order terms in $v_n$, so that 
the values of $v_1$ and $v_2$ obtained by the above method 
may have systematic errors of relative order $v_1$ and $v_2$:
for instance, if the elliptic flow is $v_2=5\%$, one may 
find instead $5.25\%$ due to this effect.

Finally, it should be noted that all corrections for acceptance 
inhomogeneities, in Eqs.~(\ref{c24accv1}) and (\ref{c24accv2}), 
involve at least the squared norm $|a_p|^2$ of the acceptance Fourier 
coefficients. 
Therefore, we believe that if all $a_{p>0}$ are smaller (in norm) 
than 0.1, then the procedure exposed in this subsection is 
superfluous: the relative magnitudes of the corresponding acceptance 
corrections are at most of a few percent, much too small to be 
significant. 
Moreover, this also applies to differential flow, since as we shall 
see shortly, in the case of a perfect detector for integrated flow, 
there are no acceptance correction even for $v'_p$.

\subsection{Differential flow}
\label{s:accdiff}

For the protons whose differential flow is measured, we introduce two 
functions $A'(\psi)$ and $P'(\psi-\Phi_R)$ which play the same role as 
$A(\phi)$ and $P(\phi-\Phi_R)$ for the pions (please note that $A'$ 
and $P'$ are NOT the derivatives of $A$ and $P$). 
Their expansion in Fourier series reads as in Eqs.~(\ref{Aphi}) 
and (\ref{Pphi}), with $a_p$ (resp. $v_p$) replaced by $a'_p$ 
(resp. $v'_p$). 

In order to calculate the generating function of the 
cumulants, Eq.~(\ref{defcm}), we need to evaluate 
$\mean{e^{ip\psi}G_n(z)}$, where the average is taken 
over all the detected protons. 
In computing this average, one must take into account carefully that 
the probability $p(\Phi_R)$ that an emitted  proton be detected depends 
on the orientation of the reaction plane $\Phi_R$:
\begin{equation}
\label{probaproton}
p(\Phi_R) = \int_0^{2\pi}A'(\psi)P'(\psi-\Phi_R){d\psi\over 2\pi}.
\end{equation} 
With this notation, we may write 
\begin{equation}
\label{inter1}
\mean{e^{ip\psi}G_n(z)}=
\frac{\int_0^{2\pi}\mean{e^{ip\psi}G_n(z)|\Phi_R}p(\Phi_R)d\Phi_R}
{\int_0^{2\pi}p(\Phi_R)d\Phi_R}.
\end{equation}
The denominator is equal to $2\pi a'_0=2\pi$ since we normalize the 
acceptance function by $a'_0=1$. 
In order to evaluate the numerator, we neglect nonflow correlations. 
Then, the proton and the pions are emitted independently 
for a fixed $\Phi_R$, so that averages factorize: 
\begin{equation}
\label{inter2}
\mean{e^{ip\psi}G_n(z)|\Phi_R} = \mean{e^{ip\psi}|\Phi_R}\mean{G_n(z)|\Phi_R}. 
\end{equation}
In this expression, $\mean{G_n(z)|\Phi_R}$ is given by Eq.~(\ref{genfunc}) 
and $\mean{e^{ip\psi}|\Phi_R}$ by an 
equation similar to (\ref{eq3}), with $a_p$ (resp. $v_p$) replaced 
by $a'_p$ (resp. $v'_p$). Using Eq.~(\ref{probaproton}), we may 
thus write 
\begin{equation}
\label{eqdif1}
\mean{e^{ip\psi}|\Phi_R}p(\Phi_R) =
\int_0^{2\pi}  e^{ip\psi} A'(\psi) P'(\psi){d\psi\over 2\pi}=
\sum_{q=-\infty}^{+\infty} a'_{q-p} v'_q e^{iq\Phi_R}.
\end{equation}
Inserting Eqs.~(\ref{inter2}) and (\ref{eqdif1}) in expression 
(\ref{inter1}), the generating function of the cumulants, 
Eq.~(\ref{defcm}), reads 
\begin{equation}
{\cal D}_{p/n}(z)=
\sum_{q=-\infty}^{+\infty} a'_{q-p} v'_q 
\frac{\int_0^{2\pi}e^{iq\Phi_R}\mean{G_n(z)|\Phi_R}d\Phi_R}
{\int_0^{2\pi}\mean{G_n(z)|\Phi_R}d\Phi_R}.
\end{equation}
Comparing with Eq.~(\ref{cpflow}), to which this equation reduces when 
the acceptance is perfect, one sees that the cumulants involve in 
general all harmonics $v'_q$.

Expanding in powers of $z$ and $z^*$ and performing the 
integrals over $\Phi_R$, one finally obtains for differential 
directed flow 
\begin{mathletters}
\label{d24accv'1}
\begin{eqnarray}
d_{1/1}\{2\}&=& 
{\Re}\left[ 1-|a_1|^2 + (a'_2)^{*} (a_2-a_1^2)\right] v'_1v_1\cr
& & + 
{\Re}\left[ a'_1 (a_1^*-a_2^* a_1)+(a'_3)^* (a_3-a_1a_2)\right]v'_2 v_2 ,
\end{eqnarray}
where $\Re$ means that one must take the real part: when the 
acceptance is not perfect, the coefficient is in general complex. 
The higher order cumulant is given by 
\begin{eqnarray}
d_{1/1}\{4\}&=& -{\Re}\left[ (1-|a_1|^2)\left((1-|a_1|^2)^2 +
2 \left|a_2-a_1^2\right|^2\right)
+(a'_2)^* (a_2-a_1^2)\left(2(1-|a_1|^2)^2 +
\left|a_2-a_1^2\right|^2\right)\right] v'_1v_1^3\cr
& & - 
{\Re}\left[ a'_1 (a_1^*-a_2^* a_1)\left( \left|a_1^*-a_2^* a_1\right|^2+
2\left|a_3-a_1a_2\right|^2\right) \right. \cr
& & \qquad \left.
+(a'_3)^* (a_3-a_1a_2)\left(2\left|a_1^*-a_2^* a_1\right|^2+
\left|a_3-a_1a_2\right|^2\right)\right]v'_2 v_2^3 
\end{eqnarray}
\end{mathletters}
Differential elliptic flow measured with respect to integrated directed 
flow yields 
\begin{eqnarray}
\label{d3accv'1}
d_{2/1}\{3\}&=&
{\Re}\left[ (1-|a_1|^2)^2+(a'_4)^*
(a_2-a_1^2)^2\right]v'_2v_1^2\cr
&+&2
{\Re}\left[ (a'_1)^* (a_1^*-a_2^* a_1) (a_2-a_1^2)
+(a'_3)^* (a_3-a_2a_1)(1-|a_1|^2)\right] \,v'_1v_1v_2.
\end{eqnarray}
The differential elliptic flow with respect to integrated elliptic
flow is given by 
\begin{mathletters}
\label{d24accv'2}
\begin{eqnarray}
\label{d2accv'2}
d_{2/2}\{2\}&=& 
{\Re}\left[ 1-|a_2|^2 + (a'_4)^{*} (a_4-a_2^2)\right] v'_2v_2\cr
& & + 
{\Re}\left[ (a'_1)^* (a_1-a_2 a_1^*)+(a'_3)^*
(a_3-a_1a_2)\right]v'_1 v_1
\end{eqnarray}
and to higher order 
\begin{eqnarray}
\label{d4accv'2}
d_{2/2}\{4\}&=& 
-{\Re}\left[ (1-|a_2|^2)\left((1-|a_2|^2)^2 +
2 \left|a_4-a_2^2\right|^2\right)
+(a'_4)^* (a_4-a_2^2)\left(2(1-|a_2|^2)^2 +
\left|a_4-a_2^2\right|^2\right)\right] v'_2v_2^3\cr
& &  -
{\Re}\left[ (a'_1)^* (a_1-a_2 a_1^*)\left( \left|a_1^*-a_2^* a_1\right|^2+
2\left|a_3-a_1a_2\right|^2\right) \right. \cr
 & & \qquad \left. 
+(a'_3)^* (a_3-a_1a_2)\left(2\left|a_1^*-a_2^* a_1\right|^2+
\left|a_3-a_1a_2\right|^2\right)\right] v'_1 v_1^3 
\end{eqnarray}
\end{mathletters}
When the acceptance of the detector used for the measurement 
of integrated flow is perfect, i.e. $a_p=0$ for $p\not= 0$, 
these formulas reduce to Eqs.~(\ref{m=1}) and (\ref{m=2}), i.e., they 
do not depend on the differential acceptance coefficients $a'_p$.

\section{Statistical errors}
\label{s:statistical}

In this appendix, we calculate the statistical fluctuations of the 
reconstructed integrated and differential flows, due to the finite 
number of events $\N$. 
More precisely, we calculate the covariance matrices
$\mean{v_n\{2k\}v_n\{2 l\}}-
\mean{v_n\{2k\}}\mean{v_n\{ 2l\}}$ and
$\mean{v'_{mn/n}\{2k+m+1\}v'_{mn/n}\{2l+m+1\}}-
\mean{v'_{mn/n}\{2k+m+1\}}\mean{v'_{mn/n}\{2l+m+1\}}$
which contain the standard error on each estimate  $v_n\{2k\}$ and 
$v'_{mn/n}\{2k+m+1\}$, and also the linear correlation between estimates 
of different orders. 
Throughout the appendix, we neglect nonflow correlations and assume 
that the detector is perfect. 

We first introduce some notations. If $x$ is an observable measured in an 
event (multiplicity, transverse energy, etc.), we denote by $\sample{x}$ 
the average value of $x$ over the available sample of events, which we 
also call the {\it sampling average\/}:
\begin{equation}
\label{sampling}
\sample{x}={1\over\N}\sum_{\alpha=1}^{\N} x_\alpha.
\end{equation}
The exact statistical average, corresponding to the limit $\N\to\infty$, 
will be denoted by $\mean{x}$. Note that $\mean{\sample{x}} = \mean{x}$.

If $y$ denotes another observable associated with each event, then 
the covariance of the sampling averages $\sample{x}$ and 
$\sample{y}$ is 
\begin{equation}
\label{samplingcov}
\mean{\sample{x}\sample{y}}-\mean{x}\mean{y}
={1\over\N}\left(\mean{xy}-\mean{x}\mean{y}\right),
\end{equation}
where we have used the property that events are statistically 
independent. 
When $y=x$, the square root of the r.-h.~s. gives the 
standard deviation of $\sample{x}$ from the exact average
$\mean{x}$, which scales like $1/\sqrt{\N}$. This result 
will prove useful later on. 

\subsection{Two-point correlation function}
\label{s:twopoint}

As explained in Sec.~\ref{s:int_gene}, one constructs for each event a 
generating function $G_n(z)$, defined by Eq.~(\ref{newG0}). 
The generating function of azimuthal correlations is the statistical 
average of this function, $\mean{G_n(z)}$. 
This quantity, given by Eq.~(\ref{meanG01}), does not depend on the 
phase of $z$ because of isotropy.
On the other hand, one measures experimentally a sampling average 
$\sample{G_n(z)}$, which generally has a (weak) dependence on 
the phase of $z$, due to statistical fluctuations. 

In the following, we shall need to evaluate the statistical fluctuations 
of the sampling average $\sample{G_n(z)}$ around the true statistical average 
$\mean{G_n(z)}$. These statistical fluctuations are characterized 
by the two-point correlation function
\begin{equation}
\label{twopointsampling}
\mean{\sample{G_n(z)}\sample{G_n(z')}} -
\mean{G_n(z)}\mean{G_n(z')}=
{1\over \N}\left( 
\mean{ G_n(z)G_n(z')} - 
\mean{ G_n(z)}\mean{G_n(z')}\right),
\end{equation}
where we have used Eq.~(\ref{samplingcov}). 
To evaluate the r.-h.\ s. of Eq.~(\ref{twopointsampling}), we first 
perform the average for a fixed orientation of the reaction plane $\Phi_R$. 
Using Eqs.~(\ref{newG0}) and (\ref{vnphir}), we obtain
\begin{eqnarray}
\label{twopointG00}
\mean{ G_n(z)G_n(z')|\Phi_R} &=&
\left( 1+{(z+z')v_n\, e^{-in\Phi_R}+{\rm c.c.}\over M}
+{z^*z'+{\rm c.c.}\over M^2}\right)^M\cr
&\simeq &
\exp\left((z+z')v_n\, e^{-in\Phi_R}+{z^*z'\over M}+{\rm c.c.}\right)
\end{eqnarray}
where ${\rm c.c.}$ denotes the complex conjugate. 
One must then average over $\Phi_R$:
\begin{equation}
\label{twopointG0}
\mean{ G_n(z)G_n(z')}=\int_0^{2\pi} {d\Phi_R\over 2\pi}
\mean{ G_n(z)G_n(z')|\Phi_R}
\end{equation}
This function is invariant under a global rotation 
$(z,z')\to (ze^{i\alpha},z'e^{i\alpha})$, but  
depends on the relative phase $e^{i\theta}\equiv z^*z'/|zz'|$ 
between $z'$ and $z$. 
We can expand it in Fourier series with respect to $\theta$, 
in the form 
\begin{equation}
\label{twopointfourier}
\mean{ G_n(z)G_n(z')} =
\sum_{m=-\infty}^{+\infty}{\cal G}_m(|z|,|z'|) 
\left({z^*z'\over |zz'|}\right)^m,
\end{equation} 
where the Fourier coefficients are given by 
\begin{eqnarray}
\label{defim}
{\cal G}_m(|z|,|z'|) 
&\equiv&\int_0^{2\pi}{d\theta\over 2\pi} e^{-im\theta}
\mean{ G_n(|z|\,e^{i\theta})G_n(|z'|)}\cr
&=&
\int_0^{2\pi} {d\theta\over 2\pi} e^{-im\theta}
\int_0^{2\pi} {d\Phi_R\over 2\pi}
\exp\left(|z|v_n\, e^{i(\theta-n\Phi_R)}+|z'|v_n\,e^{-in\Phi_R}
+{|zz'|\over M}e^{i\theta}+{\rm c.c.}\right).
\end{eqnarray}
If flow is small, more precisely if $v_n\ll 1/\sqrt{M}$, we can set 
$v_n=0$ in this equation. Then, the integral over $\Phi_R$ is trivial, 
while the integral over $\theta$ yields 
\begin{equation}
\label{solnoflow}
{\cal G}_m(|z|,|z'|) =I_m\left({2|zz'|\over M}\right).
\end{equation}
When flow is larger, the integrations can be performed using the 
following identity, valid for real $x$: 
\begin{equation}
\label{bateman}
\exp\left(x\,e^{i\phi}+x\, e^{-i\phi}\right)=
\sum_{q=-\infty}^{+\infty} e^{iq\phi} I_q(2 x). 
\end{equation}
Note that $I_{-q}(2x)=I_q(2 x)$. 
Applying Eq.~(\ref{bateman}) to the three terms in the exponential 
in Eq.~(\ref{defim}), one obtains 
\begin{equation}
\label{solcompacte}
{\cal G}_m(|z|,|z'|)= \sum_{q=-\infty}^{+\infty}
I_q(2|z|v_n)I_q(2|z'|v_n)I_{q+m}\left({2|zz'|\over M}\right). 
\end{equation}
When $v_n=0$, all terms in the sum vanish but $q=0$, and one recovers 
Eq.~(\ref{solnoflow}).

\subsection{Integrated flow}
\label{s:stat_int}

Experimentally, the generating function of the cumulants 
${\cal C}_n(z)$ is obtained by replacing $\mean{G_n(z)}$ with 
the sampling average $\sample{G_n(z)}$ in Eq.~(\ref{defc0}). 
The cumulants $c_n\{2k\}$ used to estimate the integrated 
flow are then obtained through a power-series expansion. 
We first introduce the notation 
\begin{equation}
\label{defcoef}
\left. f(z)\right|_{z^k z^{*l}}\equiv \left.{\partial^k\over\partial z^k}
{\partial^l\over\partial z^{*l}} f(z)\right|_{z=0}.
\end{equation}
With this notation, the cumulant $c_n\{2k\}$, defined by 
Eqs.~(\ref{defc1}) and (\ref{defc2k}), can be written 
\begin{equation}
\label{autredefc2k}
c_n\{2k\}=\left.{\cal C}_n(z)\right|_{z^kz^{*k}}.
\end{equation}
The statistical fluctuations of the cumulants $c_n\{2k\}$ 
is characterized by their covariance matrix:
\begin{equation}
\label{covcumul}
\mean{c_n\{2k\}c_n\{2l\}}-\mean{c_n\{2k\}}\mean{c_n\{2l\}}
=\left.\mean{{\cal C}_n(z)\,{\cal C}_n(z')}
-\mean{{\cal C}_n(z)}\mean{{\cal C}_n(z')}
\right|_{|z|^{2k}|z'|^{2l}}
\end{equation}
Let us evaluate the r.-h.~s of this equation. 
We assume that the multiplicity is large enough, so that 
${\cal C}_n(z)$, defined through Eq.~(\ref{defc0}), reduces 
to $\ln\sample{G_n(z)}$ for the sample of events. 
Expanding the logarithm to first order around $\mean{G_n(z)}$, one obtains:
\begin{eqnarray}
\label{covlogG0}
\mean{{\cal C}_n(z)\,{\cal C}_n(z')}
-\mean{{\cal C}_n(z)}\mean{{\cal C}_n(z')}
&=&
{\mean{\sample{G_n(z)}\,\sample{G_n(z')}}
-\mean{G_n(z)}\mean{G_n(z')}
\over\mean{G_n(z)}\mean{G_n(z')}}\cr
&=&{1\over \N}\left( 
{\mean{ G_n(z)G_n(z')} \over
\mean{ G_n(z)}\mean{G_n(z')}
}-1\right)
\end{eqnarray}
where we have used Eq.~(\ref{twopointsampling}). 
According to Eq.~(\ref{covcumul}), we must isolate the terms proportional 
to $|z|^{2k}|z'|^{2l}$ in the power series expansion, i.e., terms which 
do not depend on the phases of $z$ and $z'$. 
The denominator, given by Eq.~(\ref{meanG01}), depends only on $|z|$ 
and $|z'|$. 
On the other hand, the numerator depends on the relative phase of $z$ 
and $z'$. 
The isotropic part is the term $m=0$ in the Fourier expansion 
(\ref{twopointfourier}). 
We thus obtain:
\begin{equation}
\label{solcompacte2}
\mean{c_n\{2k\}c_n\{2l\}}-\mean{c_n\{2k\}}\mean{c_n\{2l\}}
 = \left.{1\over \N}\left(
{\displaystyle {\cal G}_{0}(|z|,|z'|)\over
I_0(2|z|v_n)I_0(2|z'|v_n)}-1\right)
\right|_{|z|^{2k}|z'|^{2l}}
\end{equation}
where ${\cal G}_0(|z|,|z'|)$ is given by Eq.~(\ref{solcompacte})
with $m=0$. 

The estimate of integrated flow $v_n\{2k\}$ 
is obtained by expanding Eq.~(\ref{cflow}) to order $|z|^{2k}$. 
Using Eq.~(\ref{autredefc2k}), it is related to $c_n\{2k\}$ by 
\begin{equation}
\label{defvk}
c_n\{2k\}=
\left.\ln I_0(2|z|v_n\{2k\})\right|_{z^kz^{*k}}.
\end{equation}
Using this equation, one easily relates the covariance matrix 
of the estimates $v_n\{2k\}$ to that of the corresponding 
cumulants. 
Noting that 
\begin{equation}
\label{deltav}
\ln I_0(2|z|(v_n+\delta v))-\ln I_0(2|z|v_n)= 
2\,|z|\, {I_1(2|z|v_n)\over I_0(2|z|v_n)}\,\delta v, 
\end{equation} 
we obtain from Eq.~(\ref{defvk})
\begin{equation}
\label{covv2k}
\mean{c_n\{2k\}c_n\{2l\}}-\mean{c_n\{2k\}}\mean{c_n\{2l\}}
 =  \left. 4|zz'|{I_1(2|z|v_n)\over I_0(2|z|v_n)}{I_1(2|z'|v_n)\over
I_0(2|z'|v_n)}\right|_{|z|^{2k}|z'|^{2l}}
\left(\mean{v_n\{2k\}v_n\{ 2l\}}-\mean{v_n\{2k\}}
\mean{v_n\{ 2l\}}\right).
\end{equation}
Using the expression of the covariance matrix 
of the cumulants obtained above, Eq.~(\ref{solcompacte2}), one thus 
obtains a compact expression for the covariance matrix of 
the estimates $v_n\{2k\}$. 

Before giving explicit results for the lowest order cumulants, 
let us discuss the weak flow and strong flow limits. 
If $v_n\ll 1/\sqrt{M}$, ${\cal G}_0(|z|,|z'|)$ reduces to 
Eq.~(\ref{solnoflow}), and the r.-h.~s of Eq.~(\ref{solcompacte2}) 
depends only on $|zz'|$, thus terms with 
$k\not=l$ vanish: correlations between different cumulants 
vanish in this limit, so that different estimates of $v_n$ 
are uncorrelated. 
If $v_n\gg 1/\sqrt{M}$, one can expand Eq.~(\ref{solcompacte}) to 
order $1/M$, which yields
\begin{equation}
{\cal G}_0(|z|,|z'|)=I_0(2|z|v_n)I_0(2|z'|v_n)+
{2\,|zz'|\over M}I_1(2|z|v_n)I_1(2|z'|v_n).
\end{equation}
Inserting this expression into Eq.~(\ref{solcompacte2}), and
comparing with Eq.~(\ref{covv2k}), we obtain
\begin{equation}
\label{grandv}
\mean{v_n\{2k\}v_n\{ 2l\}}-\mean{v_n\{2k\}}
\mean{v_n\{ 2l\}}={1\over 2M\N}. 
\end{equation}
In this limit, all estimates $v_n\{2k\}$ coincide and the error on 
the integrated flow is $1/\sqrt{2M\N}$, independent of $k$.
This result can be easily understood: when $v_n$ is large compared
to $1/\sqrt{M}$, the reaction plane $\Phi_R$ can be reconstructed with 
very good accuracy. Then, the integrated flow $v_n\{2k\}$ can be 
obtained as the average over all particles of all events of
$\cos [n(\phi_j-\Phi_R)]$. Since the total number of particles 
is $M\N$, one evaluates $(1/M\N)\sum_{j=1}^{M\N} \cos[n(\phi_j-\Phi_R)]$, 
and the average value of the square of this quantity is 
$1/(2M\N)$ for random angles. 
Thus the resulting statistical error is $1/\sqrt{2M\N}$. 

Finally, in more general case when $v_n$ and $1/\sqrt{M}$ are of the 
same order of magnitude, explicit expressions for the lowest order 
estimates are obtained by expanding Eqs.~(\ref{solcompacte2}) 
and (\ref{covv2k}) in power 
series of $|z|^2$ and $|z'|^2$. The standard deviations on the first 
order estimates are given by 
\begin{eqnarray}
\label{erstatint}
\left(\delta v_n\{2\}\right)^2\equiv
\mean{v_n\{2\}^2}-\mean{v_n\{2\}}^2 
&=&{1\over 2M\N}\,{1+2\, \chi^2\over 2\chi^2}\cr
\left(\delta v_n\{4\}\right)^2\equiv
\mean{v_n\{4\}^2}-\mean{v_n\{4\}}^2 
&=&{1\over 2M\N}\,{1+4\, \chi^2+\chi^4+2\chi^6\over 2\chi^6}\cr
\left(\delta v_n\{6\}\right)^2\equiv
\mean{v_n\{6\}^2}-\mean{v_n\{6\}}^2 
&=&{1\over 2M\N}\,{3+18\, \chi^2+9\chi^4+28\chi^6+12\chi^8
+24\chi^{10}\over 24\chi^{10}}
\end{eqnarray}
with $\chi^2\equiv Mv_n^2$. 
In the limit $\chi\gg 1$, they results reduce to Eq.~(\ref{grandv}). 

\begin{center}
\begin{figure}[ht!]
\centerline{\includegraphics*[width=0.6\linewidth]{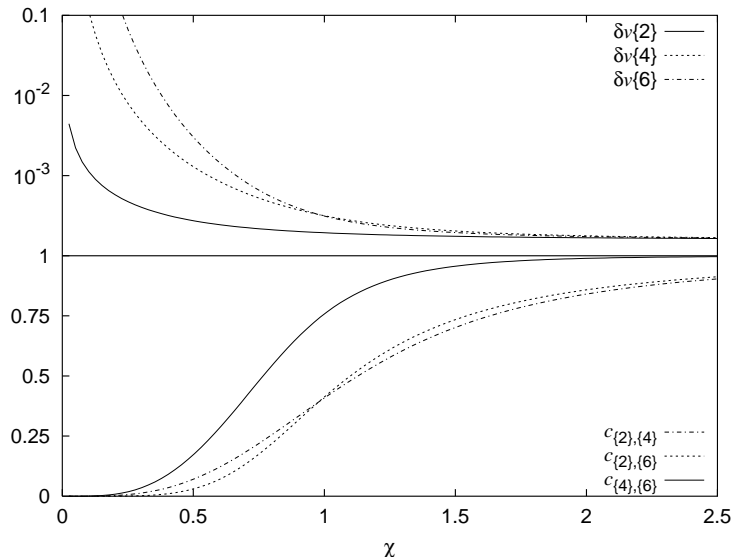}}
\medskip
\caption{Top: statistical errors on $v_n\{2\}$, $v_n\{4\}$ and $v_n\{6\}$
for $\N=10^5$ and $M=200$, as a function of $\chi\equiv v_n\sqrt{M}$. 
Bottom: linear correlation $c_{\{2k\},\{2l\}}$ between each pair of 
estimates $v_n\{2k\}$ and $v_n\{2l\}$, with $1\leq k \leq 3$ and 
 $1\leq l \leq 3$.}
\label{fig:stat}
\end{figure}
\end{center}

The values of $\delta v_n\{2k\}$ given by Eqs.~(\ref{erstatint}) are 
plotted in Fig.\ \ref{fig:stat} as a function of $\chi$. 
We have taken $\N=10^5$ events, with multiplicity $M=200$ each. 
With these values, $\chi=1$ corresponds to $v_n\simeq 7\%$. 
Some comments on these results:
\begin{itemize}
\item 
The statistical error increases rapidly as $v_n$ decreases: 
if $v_n=3\%$ (i.e., $\chi=0.42$), the statistical error using the 
fourth-order cumulant is reasonably small, less than $0.2\%$. 
But if $v_n=1.5\%$ ($\chi=0.21$) the statistical error becomes of the 
same magnitude as the flow itself. 
Such small values of the flow can be studied only with a higher 
multiplicity and/or a very large number of events. 
\item 
While the lowest order estimate $v_n\{2\}$ always has the smallest 
statistical error, the statistical error on $v_n\{6\}$ is slightly 
smaller (by at most 5\%) than the error on $v_n\{4\}$ for $\chi>1$.
\end{itemize}

The correlation between estimates of different orders is given by 
\begin{eqnarray}
\label{corint}
\mean{v_n\{2\}v_n\{4\}}-\mean{v_n\{2\}}\mean{v_n\{4\}}
&=&{1\over 2M\N}\cr
\mean{v_n\{2\}v_n\{6\}}-\mean{v_n\{2\}}\mean{v_n\{6\}}
&=&{1\over 2M\N}\cr
\mean{v_n\{4\}v_n\{6\}}-\mean{v_n\{4\}}\mean{v_n\{6\}}
&=&{1\over 2M\N}\,{3\chi^4+\chi^6+2\chi^8\over 2\chi^8}
\end{eqnarray}
Figure \ref{fig:stat} displays the linear correlation $c_{\{2k\},\{2l\}}$ 
between $v_n\{2k\}$ and $v_n\{2l\}$ with $k\not=l$. We recall that the 
linear correlation $c$ between two random variables $x$ and $y$ is defined 
as 
\begin{equation}
c\equiv \frac{\langle xy\rangle-\langle x\rangle\langle y\rangle}
{\sqrt{\langle x^2\rangle-\langle x\rangle^2}
\sqrt{\langle y^2\rangle-\langle y\rangle^2}}.
\end{equation}
$c$ always lies between $-1$ and $1$; these two limiting cases 
corresponding to a linear relation $y=a x$, while $c=0$ if $x$ and  $y$ 
are uncorrelated. 
As expected from the discussion following Eq.~(\ref{grandv}), 
different estimates are uncorrelated if $\chi\ll 1$, but 
the correlation becomes stronger and stronger as the resolution $\chi$ 
increases. 

\begin{center}
\begin{figure}[ht!]
\centerline{\includegraphics*[width=0.5\linewidth]{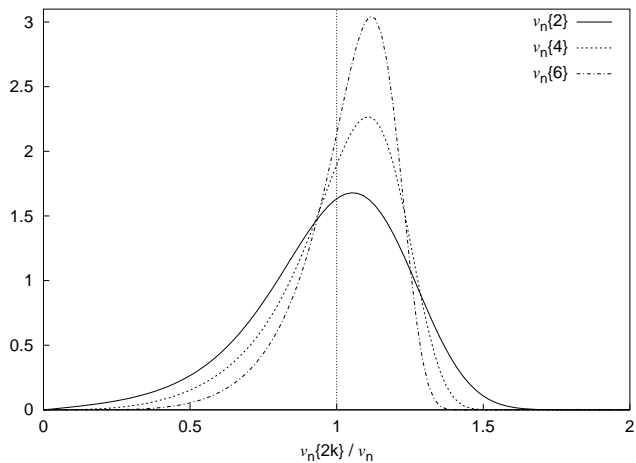}}
\medskip
\caption{Distribution of $v_n\{2\}$, $v_n\{4\}$ and $v_n\{6\}$, scaled 
by the theoretical value $v_n$.}
\label{fig:nongaussian}
\end{figure}
\end{center}

Finally, we would like to point out that the statistical fluctuations 
of the estimate $v_n\{2k\}$ around the true value $v_n$ are not 
Gaussian. 
Indeed, it can be shown that the fluctuations of the cumulants are 
generally Gaussian.
Thus, according to Eqs.~(\ref{c246}), the fluctuations of the variable 
$\xi\equiv v_n\{2k\}^{2k}$ are Gaussian, but not the fluctuations of 
$v_n\{2k\}$ itself. We may write 
\begin{equation}
\label{discumulant}
\frac{dN}{d\xi}={1\over\sqrt{2\pi\sigma}} \exp\left(-
\frac{(\xi-v_n^{2k})^2}{2\sigma^2}\right),
\end{equation}
where $\sigma$ is the standard deviation on $\xi$. 
It is related to the deviation $\delta v_n\{2k\}$, 
Eqs.~(\ref{erstatint}), by 
\begin{equation}
\label{sigcumulant}
\sigma=2\, k\, v_n^{2k-1}\,\delta v_n\{2k\}.
\end{equation}
In order to illustrate the non-Gaussian character of the fluctuations of 
$v_n\{2k\}$, we display in Fig.\ \ref{fig:nongaussian} the distribution of 
$v_n\{2\}$, $v_n\{4\}$ and $v_n\{6\}$ when the average value of $\xi$ in 
Eq.~(\ref{sigcumulant}) is only one standard deviation above zero, i.e., 
when $\sigma=v_n^{2k}$. 
Then, using Eq.~(\ref{discumulant}), the probability that $\xi<0$ is 
about 16\%, in which case $v_n\{2k\}$ is undefined. 
One notes that for the 84\% remaining cases, the distribution of $v_n\{2k\}$ 
becomes narrower as $2k$ increases. 

\subsection{Differential flow}
\label{s:stat_diff}

The generating function of the cumulants used for differential flow, 
${\cal D}_{p/n}(z)$, is the ratio of two quantities:
the numerator of the r.-h.~s. of Eq.~(\ref{defcm}) 
is evaluated from a sample of $\Np$ protons, 
while the denominator is 
calculated from the same sample of events as for the integrated flow, 
that is, $\N$.
Since the measurement of differential flow is usually performed in a 
narrow phase-space window, we neglect the contribution of the denominator 
to the statistical error. We thus write 
\begin{equation}
\label{defcmsample}
{\cal D}_{p/n}(z)=\frac{\sample{e^{ip\psi}G_n(z)}}{\mean{G_n(z)}},
\end{equation}
where the denominator is given by Eq.~(\ref{meanG01}). 

The cumulants $d_{mn/n}\{2k+m+1\}$ defined by Eqs.~(\ref{defcm1}) and 
(\ref{defdk}) can be written in the form 
\begin{equation}
\label{autredefdk}
d_{mn/n}\{2k+m+1\}= 
\left. {1\over 2}\left({\cal D}_{mn/n}(z)+{\cal D}^*_{mn/n}(z^*)\right)
\right|_{z^{k+m}z^{*k}}.
\end{equation}
We now evaluate their  covariance matrix. 
For simplicity, we shall assume that there is no differential flow, 
i.e., that the $\psi$ distribution is isotropic, so that 
$\mean{{\cal D}_{p/n}(z)}=0$, and only fluctuations remain. 
Using Eq.~(\ref{samplingcov}) and the definition (\ref{defcmsample}), 
we thus obtain  
\begin{eqnarray}
\mean{{\cal D}_{p/n}(z){\cal D}_{p/n}(z')}&=&0\cr
\mean{{\cal D}_{p/n}(z){\cal D}^*_{p/n}(z')}&=&{1\over\Np}
\frac{\mean{G_n(z)G_n(z')}}{\mean{G_n(z)}\mean{G_n(z')}}.
\end{eqnarray}
We recognize in the last equation the two-point
function studied in Appendix \ref{s:twopoint}. 
The covariance matrix of the cumulants (\ref{autredefdk})
can thus be written 
\begin{eqnarray}
\label{covcumuldif}
\mean{d_{mn/n}\{2k+m+1\}d_{mn/n}\{2l+m+1\}}
&=& 
\left. {1\over 4\Np}\left(
\frac{\mean{G_n(z)G_n(z'^*)}}{\mean{G_n(z)}\mean{G_n(z'^*)}}+
\frac{\mean{G_n(z^*)G_n(z')}}{\mean{G_n(z^*)}\mean{G_n(z')}}\right)
\right|_{z^{k+m}z^{*k}z'^{l+m}z'^{*l}}\cr
&=&\left. {1\over 2\Np}
\frac{\mean{G_n(z^*)G_n(z')}}{\mean{G_n(z)}\mean{G_n(z')}}
\right|_{z^{k+m}z^{*k}z'^{l+m}z'^{*l}}.
\end{eqnarray}
Using Eq.~(\ref{meanG01}) and 
the Fourier decomposition of the two-point function 
(\ref{twopointfourier}), we obtain 
\begin{equation}
\label{covcumuldif1}
\mean{d_{mn/n}\{2k+m+1\}d_{mn/n}\{2l+m+1\}}
=\left. {1\over 2\Np}
\frac{{\cal G}_m(|z|,|z'|)}{I_0(2|z|v_n)I_0(2|z'|v_n)}
\left(\frac{zz'}{|zz'|}\right)^m
\right|_{z^{k+m}z^{*k}z'^{l+m}z'^{*l}}.
\end{equation}
where ${\cal G}_m(|z|,|z'|)$ is given by Eq.~(\ref{solcompacte}). 

The estimate of differential flow $v'_{mn/n}\{2k+m+1\}$ 
is obtained by expanding Eq.~(\ref{cpflow}) to order $z^{k+m}z^{*k}$. 
Using Eq.~(\ref{autredefdk}), it is related to 
$d_{mn/n}\{2k+m+1\}$ by 
\begin{equation}
\label{samplingdiff}
d_{mn/n}\{2k+m+1\}=
\left.{I_m(2|z|v_n)\over I_0(2|z|v_n)}
\left(\frac{z}{|z|}\right)^m
\right|_{z^{k+m}z^{*k}}
\, v'_{mn/n}\{2k+m+1\}
\end{equation}
Using this equation, one easily relates the covariance matrix 
of the estimates $v'_{mn/n}\{2k+m+1\}$ to that of the corresponding 
cumulants, given by Eq.~(\ref{covcumuldif1}): 
\begin{equation}
\label{soldifflow}
\left.{I_m(2|z|v_n)\over I_0(2|z|v_n)}
{I_m(2|z'|v_n)\over I_0(2|z'|v_n)}\right|_{|z|^{2k}|z'|^{2l}}
\mean{v'_{mn/n}\{2k+m+1\}v'_{mn/n}\{2l+m+1\}}
={1\over 2\Np}
\left.{{\cal G}_m(|z|,|z'|)\over I_0(2|z|v_n) I_0(2 |z'|v_n)}
\right|_{|z|^{2k}|z'|^{2l}}.
\end{equation}

Let us discuss the weak flow and strong flow limits.
When $v_n \ll 1/\sqrt{M}$, 
${\cal G}_m(|z|,|z'|)$ is given by 
Eq.~(\ref{solnoflow}), and terms with $k\not= l$ vanish: as in the 
case of integrated flow, correlations between estimates of different 
orders vanish in this limit. 
When $v_n\gg 1/\sqrt{M}$, expanding Eq.~(\ref{solcompacte}) to 
leading order in $1/M$, we obtain
\begin{equation}
{\cal G}_m(|z|,|z'|)=I_m(2|z|v_n)I_m(2|z'|v_n).
\end{equation}
Then, the covariance matrix, Eq.~(\ref{soldifflow}), reduces to
\begin{equation}
\label{difflargev}
\mean{v'_{mn/n}\{2k+m+1\}v'_{mn/n}\{2l+m+1\}}
-\mean{v'_{mn/n}\{2k+m+1\}}\mean{v'_{mn/n}\{2l+m+1\}}
={1\over 2\Np}.
\end{equation}
This result can be simply understood by repeating the argument 
used in the case of integrated flow. 

\begin{center}
\begin{figure}[ht!]
\centerline{\includegraphics*[width=0.62\linewidth]{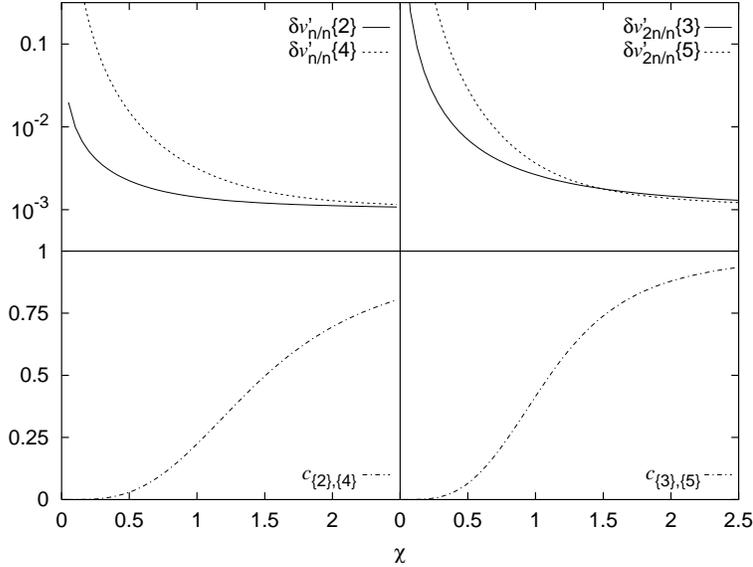}}
\medskip
\caption{Left (resp. right): statistical properties of the estimates 
of $v'_{n/n}$ (resp. $v'_{2n/n}$). 
Top: statistical errors on $v'_{n/n}\{2\}$ and $v'_{n/n}\{4\}$ 
(resp. $v'_{2n/n}\{3\}$ and $v'_{2n/n}\{5\}$) for $\Np=5\times 10^5$ 
and $M=200$, as a function of $\chi\equiv v_n\sqrt{M}$. 
Bottom: linear correlation between $v'_{n/n}\{2\}$ and $v'_{n/n}\{4\}$
(resp. $v'_{2n/n}\{3\}$ and $v'_{2n/n}\{5\}$). }
\label{fig:statd}
\end{figure}
\end{center}

In the more general case when $v_n$ and $1/\sqrt{M}$ are of the 
same order of magnitude, the following lowest order formulas
are derived by expanding Eq.~(\ref{soldifflow}) in power series. 
For $m=1$:
\begin{eqnarray}
\label{statm=1}
\mean{v'_{n/n}\{2\}^2} -\mean{v'_{n/n}\{2\}}^2&=&
{1\over 2\Np}\,{1+\chi^2\over \chi^2}\cr
\mean{v'_{n/n}\{2\}v'_{n/n}\{4\}} -
\mean{v'_{n/n}\{2\}}\mean{v'_{n/n}\{4\}}&=&
{1\over 2\Np}\cr
\mean{v'_{n/n}\{4\}^2} -\mean{v'_{n/n}\{4\}}^2&=&
{1\over 2\Np}\,{2+6\,\chi^2+\chi^4+\chi^6\over\chi^6}
\end{eqnarray}
with $\chi^2\equiv Mv_n^2$. 
For $m=2$:
\begin{eqnarray}
\label{statm=2}
\mean{v'_{2n/n}\{3\}^2} -\mean{v'_{2n/n}\{3\}}^2&=&
{1\over 2\Np}\,{2+4\,\chi^2+\chi^4\over \chi^4}\cr
\mean{v'_{2n/n}\{3\}v'_{2n/n}\{5\}} -
\mean{v'_{2n/n}\{3\}}\mean{v'_{2n/n}\{5\}}&=&
{1\over 2\Np}\,{3+\chi^2\over\chi^2}\cr
\mean{v'_{2n/n}\{5\}^2} -\mean{v'_{2n/n}\{5\}}^2&=&
{1\over 2\Np}\,{6+24\,\chi^2+9\,\chi^4+10\,\chi^6+4\,\chi^8
\over 4\,\chi^8}
\end{eqnarray}
When $\chi\gg 1$, these results reduce to (\ref{difflargev}). 
Figure \ref{fig:statd} displays the variation with $\chi$ of the standard 
errors on the various estimates and the linear correlation between 
cumulants of different orders. 
The behaviour is qualitatively the same as for integrated flow. 


\begin{thebibliography}{99}

\bibitem{Andronic:2001cx}
A.~Andronic {\it et al.}  [FOPI Collaboration],
Nucl.\ Phys.\ A {\bf 679}, 765 (2001)
[nucl-ex/0008007].

\bibitem{Ackermann:2001tr}
K.~H.~Ackermann {\it et al.}  [STAR Collaboration],
Phys.\ Rev.\ Lett.\ {\bf 86}, 402 (2001)
[nucl-ex/0009011];
R.~J.~Snellings  [STAR Collaboration],
nucl-ex/0104006.

\bibitem{Reisdorf:1997fx}
W.~Reisdorf and H.~G.~Ritter,
Ann.\ Rev.\ Nucl.\ Part.\ Sci.\  {\bf 47}, 663 (1997).

\bibitem{Herrmann:1999wu}
N.~Herrmann, J.~P.~Wessels and T.~Wienold,
Ann.\ Rev.\ Nucl.\ Part.\ Sci.\  {\bf 49}, 581 (1999).

\bibitem{Ollitrault:1998vz}
J.-Y.~Ollitrault,
Nucl.\ Phys.\ A {\bf 638}, 195c (1998)
[nucl-ex/9802005].

\bibitem{Sorge:1999mk}
H.~Sorge,
Phys.\ Rev.\ Lett.\ {\bf 82}, 2048 (1999)
[nucl-th/9812057].

\bibitem{Teaney:2000cw}
D.~Teaney, J.~Lauret and E.~V.~Shuryak,
nucl-th/0011058.

\bibitem{Nandi:1999vb}
B.~K.~Nandi, G.~C.~Mishra, B.~Mohanty, D.~P.~Mahapatra and T.~K.~Nayak,
Phys.\ Lett.\ B {\bf 449}, 109 (1999)
[nucl-ex/9812004].

\bibitem{Asakawa:2000dr}
M.~Asakawa, H.~Minakata and B.~Muller,
nucl-th/0011031.

\bibitem{Voloshin:2000xf}
S.~A.~Voloshin,
Phys.\ Rev.\ C {\bf 62}, 044901 (2000)
[nucl-th/0004042].

\bibitem{Voloshin:1996mc}
S.~A.~Voloshin and W.~E.~Cleland,
Phys.\ Rev.\ C {\bf 53}, 896 (1996)
[nucl-th/9509025].

\bibitem{Wiedemann:1998cr}
U.~A.~Wiedemann,
Phys.\ Rev.\ C {\bf 57}, 266 (1998)
[nucl-th/9707046].

\bibitem{Heiselberg:1999ik}
H.~Heiselberg,
Phys.\ Rev.\ Lett.\  {\bf 82}, 2052 (1999)
[nucl-th/9809077].

\bibitem{Lisa:2000xj}
M.~A.~Lisa {\it et al.}  [E895 Collaboration],
Phys.\ Lett.\ B {\bf 496}, 1 (2000)
[nucl-ex/0007022].

\bibitem{Voloshin:1996mz}
S.~Voloshin and Y.~Zhang,
Z.\ Phys.\ C {\bf 70}, 665 (1996)
[hep-ph/9407282].

\bibitem{Li:1999bh}
Bao An Li and A.~T.~Sustich,
Phys.\ Rev.\ Lett.\  {\bf 82}, 5004 (1999)
[nucl-th/9905038].

\bibitem{Danielewicz:1985hn}
P.~Danielewicz and G.~Odyniec,
Phys.\ Lett.\ B {\bf 157}, 146 (1985).

\bibitem{Poskanzer:1998yz}
A.~M.~Poskanzer and S.~A.~Voloshin,
Phys.\ Rev.\ C {\bf 58}, 1671 (1998)
[nucl-ex/9805001].

\bibitem{Ollitrault:1993ba}
J.-Y.~Ollitrault,
Phys.\ Rev.\ D {\bf 48}, 1132 (1993)
[hep-ph/9303247].

\bibitem{Ollitrault:1997di}
J.-Y.~Ollitrault,
nucl-ex/9711003.

\bibitem{Taranenko:1999yh}
A.~Taranenko {\it et al.}  [TAPS Collaboration],
nucl-ex/9910002.

\bibitem{Chkhaidze:2000gs}
L.~Chkhaidze, T.~Djobava and L.~Kharkhelauri,
Phys.\ Lett.\ B {\bf 479}, 21 (2000)
[hep-ex/9912035].

\bibitem{AbdAllah:2000qr}
N.~N.~Abd Allah,
J.\ Phys.\ Soc.\ Jap.\  {\bf 69}, 1068 (2000).

\bibitem{Simic:2001aq}
L.~Simic and J.~Milosevic,
J.\ Phys.\ G {\bf G27}, 183 (2001) [nucl-th/0106037].

\bibitem{Jain:1995cm}
P.~L.~Jain, G.~Singh and A.~Mukhopadhyay,
Phys.\ Rev.\ Lett.\  {\bf 74}, 1534 (1995).

\bibitem{Chung:2000ny}
P.~Chung {\it et al.}  [E895 Collaboration],
Phys.\ Rev.\ Lett.\  {\bf 85}, 940 (2000)
[nucl-ex/0101003].

\bibitem{Barrette:2001cb}
J.~Barrette {\it et al.}  [E877 Collaboration],
Phys.\ Rev.\ C {\bf 63}, 014902 (2001)
[nucl-ex/0007007].

\bibitem{Appelshauser:1998dg}
H.~Appelsh\"auser {\it et al.}  [NA49 Collaboration],
Phys.\ Rev.\ Lett.\ {\bf 80}, 4136 (1998)
[nucl-ex/9711001].

\bibitem{Schlagheck:1999ja}
H.~Schlagheck  [WA98 Collaboration],
Nucl.\ Phys.\ A {\bf 661}, 337 (1999)
[nucl-ex/9907005].

\bibitem{Wang:1991qh}
S.~Wang {\it et al.},
Phys.\ Rev.\ C {\bf 44}, 1091 (1991).

\bibitem{Prendergast:2000fv}
E.~P.~Prendergast {\it et al.},
Phys.\ Rev.\ C {\bf 61}, 024611 (2000).

\bibitem{Singh:1994ti}
G.~Singh and P.~L.~Jain,
Phys.\ Rev.\ C {\bf 49}, 3320 (1994).

\bibitem{Lacey:2001}
R.~A.~Lacey [PHENIX Collaboration],
nucl-ex/0105003.

\bibitem{Danielewicz:1988in}
P.~Danielewicz {\it et al.},
Phys.\ Rev.\ C {\bf 38}, 120 (1988).

\bibitem{Dinh:2000mn}
P.~M.~Dinh, N.~Borghini and J.-Y.~Ollitrault,
Phys.\ Lett.\ B {\bf 477}, 51 (2000)
[nucl-th/9912013].

\bibitem{Borghini:2000cm}
N.~Borghini, P.~M.~Dinh and J.-Y.~Ollitrault,
Phys.\ Rev.\ C {\bf 62}, 034902 (2000)
[nucl-th/0004026].

\bibitem{Aggarwal:1997iu}
M.~M.~Aggarwal {\it et al.}  [WA93 Collaboration],
Phys.\ Lett.\ B {\bf 403}, 390 (1997).

\bibitem{Raniwala:2000yz}
R.~Raniwala, S.~Raniwala and Y.~P.~Viyogi,
Phys.\ Lett.\ B {\bf 489}, 9 (2000)
[nucl-ex/0007016].

\bibitem{Danielewicz:1983we}
P.~Danielewicz and M.~Gyulassy,
Phys.\ Lett.\ B {\bf 129}, 283 (1983).

\bibitem{Gustafsson:1984ka}
H.~A.~Gustafsson {\it et al.},
Phys.\ Rev.\ Lett.\ {\bf 52}, 1590 (1984).

\bibitem{Ollitrault:1992bk}
J.-Y.~Ollitrault,
Phys.\ Rev.\ D {\bf 46}, 229 (1992).

\bibitem{Barrette:1994xr}
J.~Barrette {\it et al.}  [E877 Collaboration],
Phys.\ Rev.\ Lett.\ {\bf 73}, 2532 (1994)
[hep-ex/9405003].

\bibitem{Jiang:1992bw}
J.~Jiang {\it et al.},
Phys.\ Rev.\ Lett.\ {\bf 68}, 2739 (1992).

\bibitem{Crochet:2000fz}
P.~Crochet {\it et al.}  [FOPI collaboration],
Phys.\ Lett.\ B {\bf 486}, 6 (2000)
[nucl-ex/0006004].

\bibitem{Chung:2001je}
P.~Chung {\it et al.},
Phys.\ Rev.\ Lett.\  {\bf 86}, 2533 (2001)
[nucl-ex/0101002].

\bibitem{Barrette:2000ex}
J.~Barrette {\it et al.}  [E877 Collaboration],
Phys.\ Lett.\ B {\bf 485}, 319 (2000)
[nucl-ex/0004002].


\bibitem{Borghini:2001sa}
N.~Borghini, P.~M.~Dinh and J.-Y.~Ollitrault,
Phys.\ Rev.\ C {\bf 63}, 054906 (2001)
[nucl-th/0007063].

\bibitem{Borghini:2000iy}
N.~Borghini, P.~M.~Dinh and J.-Y.~Ollitrault,
nucl-th/0011013.


\bibitem{Snellings:2000bt}
R.~J.~M.~Snellings, H.~Sorge, S.~A.~Voloshin, F.~Q.~Wang and N.~Xu,
Phys.\ Rev.\ Lett.\  {\bf 84}, 2803 (2000)
[nucl-ex/9908001].

\bibitem{Tsang:1991}
M. B. Tsang {\it et al.}, 
Phys.\ Rev.\ {\bf C44}, 2065 (1991). 

\bibitem{Danielewicz:1994nb}
P.~Danielewicz,
Phys.\ Rev.\ {\bf C51}, 716 (1995).

\bibitem{Doss:1986eh}
K.~G.~Doss {\it et al.},
Phys.\ Rev.\ Lett.\ {\bf 57}, 302 (1986).

\bibitem{Gosset:1989cm}
J.~Gosset {\it et al.},
Phys.\ Rev.\ Lett.\ {\bf 62}, 1251 (1989).

\bibitem{Ogilvie:1989a}
C.~A.~Ogilvie {\it et al.}, 
Phys.\ Rev.\ {\bf C40}, 654 (1989).

\bibitem{Sullivan:1990}
J.-P. Sullivan {\it et al.}, 
Phys.\ Lett.\ {\bf B249}, 8 (1990).

\bibitem{Fai:1987bt}
G.~Fai, W.~M.~Zhang and M.~Gyulassy,
Phys.\ Rev.\ C {\bf 36}, 597 (1987).

\bibitem{Htun:1999nc}
M.~M.~Htun {\it et al.},
Phys.\ Rev.\ C {\bf 59}, 336 (1999)
[nucl-ex/9811013].

\bibitem{Ogilvie:1989b}
C.~A.~Ogilvie {\it et al.}, 
Phys.\ Rev.\ {\bf C40}, 2592 (1989).

\bibitem{Krofcheck:1989} 
D.~Krofcheck {\it et al.}, 
Phys.\ Rev.\ Lett.\ {\bf 63}, 2028 (1989).

\bibitem{Wilson:1992} 
W. K. Wilson, R. Lacey, C. A. Ogilvie, and G. D. Westfall, 
Phys.\ Rev.\ {\bf C45}, 738 (1992). 

\bibitem{Huovinen:2001cy}
P.~Huovinen, P.~F.~Kolb, U.~Heinz, P.~V.~Ruuskanen and S.~A.~Voloshin,
Phys.\ Lett.\ B {\bf 503}, 58 (2001)
[hep-ph/0101136].

\bibitem{vanKampen}
N.~G.~van Kampen, {\it Stochastic processes in physics and chemistry} 
(North-Holland, Amsterdam, 1981).

\bibitem{Ollitrault:1995dy}
J.~Y.~Ollitrault,
Nucl.\ Phys.\ A {\bf 590}, 561c (1995).

\end{thebibliography}
\end{document}